\documentclass[a4, 10pt,twocolumn]{article}

\usepackage[english]{babel}
\usepackage{graphicx,amsmath,amssymb,verbatim}
\graphicspath{{.}}
\usepackage{subfigure}
\usepackage[latin1]{inputenc}
\usepackage[all]{xy}

\begin{document}

\title{Faceted Ranking of Egos in Collaborative Tagging Systems}
\author{Jose I. Orlicki$^{(1,2)}$, Pablo I. Fierens$^{(2)}$ and J.  Ignacio Alvarez-Hamelin$^{(2,3)}$\\
{\small (1) Core Security Technologies, Humboldt 1967 $1^\circ$p, C1414CTU  Buenos Aires, Argentina}  \\
{\small (2) ITBA, Av. Madero 399, C1106ACD Buenos Aires, Argentina} \\
{\small (3) CONICET (Argentinian Council of Scientific and Technological
Research)}}
\date{}
\maketitle
\begin{abstract}
Multimedia uploaded content is tagged and recommended by users of collaborative systems, resulting in informal classifications also known as folksonomies.
Faceted web ranking has been proved a reasonable alternative to a single ranking which does not take into account a personalized context.
In this paper we analyze the online computation of rankings of users associated to facets made up of multiple tags.
Possible applications are user reputation evaluation (ego-ranking) and improvement of content quality in case of retrieval.
We propose a solution based on PageRank as centrality measure:
(i) a ranking for each tag is computed offline on the basis of the corresponding tag-dependent subgraph; (ii) a faceted order is generated by merging rankings corresponding to all the tags in the facet.
The fundamental assumption, validated by empirical observations, is that step (i) is scalable. We also present algorithms for part (ii) having time complexity ${\cal O}(k)$, where $k$ is the number of tags in the facet, well suited to online computation.
\end{abstract}

\section{Introduction}
\label{sec:introduction}
In collaborative tagging systems, users assign keywords or \emph{tags} to their uploaded content, or bookmarks, in order to improve future navigation, filtering or searching (see, e.g., Marlow {\em et al.}~\cite{tagging}). These systems generate a categorization of content commonly known as a \emph{folksonomy}.

An example is the collaborative URL tagging system \emph{Delicious}~\cite{delicious}, which was analyzed in depth by Golger and Huberman~\cite{folksonomies}, discovering temporal stability in the relative proportions of tags within a given tagging subject.
In this system Internet resources (URLs) are bookmarked and classified with tags by users.
Other two well-known collaborative tagging systems for multimedia content are \emph{YouTube}~\cite{youtube} (videos) and \emph{Flickr}~\cite{Flickr} (photos), which are the focus of this paper.

\emph{YouTube} and \emph{Flickr} differ from \emph{Delicious} in that the resources are uploaded by users, so that all contents bookmarked as favorites are inside the system. That is, \emph{YouTube} and \emph{Flickr} can be considered closed systems.

Users can be ranked in relation to a tag or set of tags which we call a {\em facet}. Some applications of these \emph{faceted} (i.e., tag-associated) rankings are: (i) searching for content through navigation of the best users inside a tag-facet; (ii) measuring reputation of users by listing their best rankings for different tags or tag sets.

The order or ranking can be determined by a centrality measure, such as PageRank~\cite{pagerank_original,pagerank}, in a recommendation or subscription graph. Given a facet, a straightforward solution is to compute the centrality measure based on an appropriate facet-dependent subgraph of the recommendation network. However, the online computation of the centrality measure is unfeasible because its high time complexity, even for small facets with two or three tags. Moreover, the offline computation of the centrality measure for each facet is also unfeasible because the large number of possible facets. Therefore, alternative solutions must be looked for.
A simple solution is to use a general ranking computed offline, which is then filtered online for each facet query.
Using a single ranking of web pages or users within folksonomies has the disadvantage that the best ranked ones are those having the highest centrality in a global ranking, which is facet-independent. In the information retrieval case, this implies that the returned results are ordered in a way that does not take into account the focus on the searched topic. This problem is called \emph{topic drift}~\cite{qd-pagerank}.

In this paper we propose a solution to the problem of topic drift in faceted rankings which is based on PageRank as centrality measure. Our approach follows a two-step procedure: (i) a ranking for each tag is computed offline on the basis of the corresponding tag-dependent subgraph;
(ii) a faceted order is generated by merging rankings corresponding to all the tags in the facet.

The fundamental assumption is that step (i) in this procedure can be computed with an acceptable overhead which depends on the size of the dataset. This hypothesis is validated by two empirical observations. On one hand, in the studied recommendation (tagged) graphs most of the tags are associated to very small subgraphs, while only a small number of tags have large associated subgraphs (see Section~\ref{sec:construction}). On the other hand, the mean number of tags per edge is finite and small as explained in Section~\ref{subsec:scalability}.

The problem then becomes to find a good and efficient algorithm to merge several rankings in step (ii). In Section \ref{sec:definitions_and_algorithms}, we present several alternatives. We concentrate our effort on facets that correspond to the \emph{logical conjunction} of tags (\emph{match-all-tags}-queries) because this is the most used logical combination in information retrieval (Christopher~\cite{ir_book}, Chapter 1).

The rest of the paper is organized as follows.
We discuss prior works and their limitations in Section~\ref{sec:related}.
In Section~\ref{sec:construction} we explore two real examples of tagged graphs. In particular, we analyze several important characteristics of these graphs, such as the scale-free behavior of the vertex indegree and assortativeness of the embedded recommendation network (see Section \ref{subsec:network-analysis}). The proposed algorithms are introduced in Section~\ref{sec:definitions_and_algorithms}, including an analysis of related scalability issues in Section \ref{subsec:scalability}. We discuss experimental results in Section~\ref{sec:experiments} and we conclude with some final remarks and possible directions of future work in Section \ref{sec:conclusions}.

\section{Related work}
\label{sec:related}

Theory and implementation concepts used in this work for PageRank centrality are based on the comprehensive survey of Langville and Meyer~\cite{pagerank}.
This centrality measure for directed graphs is a variation of eigenvector centrality which includes the notion of a random surfer, i.e., an imaginary surfer that, in arriving to a vertex with no out-links, jumps to a randomly chosen vertex.
The PageRank algorithm is based on the iterated multiplication of the adjacency matrix of the directed graph (modified to add the random surfer), and a vector representing the probability that a surfer is in a particular vertex. The iteration stops when each vector component does not change more than a given error $\epsilon$. Only a hundred of matrix multiplications are needed for $\epsilon=10^{-6}$ and standard parameters (see ~\cite{pagerank} for details).

Basic topic-sensitive PageRank analysis was attempted biasing the general PageRank equation to special subsets of web pages by Al-Saffar and Heileman~\cite{personalized}, and using a predefined set of categories by Haveliwala~\cite{topic} extracted from the Open Directory Project~\cite{odp}. Although encouraging results were obtained in both works, they suffer from the limitation of a fixed number of topics biasing the rankings. Another variations of personalized PageRank were augmented with weights based on usage by Eirinaki and Vazirgiannis~\cite{Eirinaki05} and on access time-length and frequency by Guo {\em et al.}~\cite{Guo07} by previous users, they built a unique PageRank vector adapted to usage but the result is not user dependent nor query dependent as we prefer.

Hotho {\em et al.}~\cite{folkrank} adapted PageRank to work on a tripartite graph of users, tags and resources corresponding to a folksonomy. They also developed a form of topic-biasing on the modified PageRank, but the generation of a faceted ranking implies a new computation of the adapted PageRank algorithm on the network for each new facet.

There has also been some work done on faceted ranking of web pages. For example, the approach of DeLong, Mane and Srivastava~\cite{concept-aware} involves the construction of a larger multigraph using the hyperlink graph with each vertex corresponding to a pair webpage-concept and each edge to a hyperlink associated with a concept.
Subgraph ideas are suggested by them:
``It might be faster to simply run PageRank on sub-graphs pertaining to each individual concept (assuming there are a small number of concepts).''
Although DeLong {\em et al.}~\cite{concept-aware} obtain good ranking results for single-keyword facets, they do not support multi-keyword queries.

Query-dependent PageRank calculation was introduced in Richarson and Domingos~\cite{qd-pagerank} to extract a weighted probability per keyword for each webpage. These probabilities are summed up to generate a query-dependent result.
They also show that this faceted ranking has, for thousands of keywords, computation and storage requirements that are only approximately $100$-$200$ times greater than that of a single query-independent PageRank.
As we show in Section~\ref{subsec:scalability}, our facet-dependent ranking algorithms have similar time complexity.

Scalability issues were also tackled by Jeh and Widom~\cite{scaling} criticizing offline computation of multiple PageRank vectors for each possible query and preferring another more efficient dynamic programming algorithm for online calculation of the faceted rankings based on offline computation of basis vectors.
They found that their algorithm scales well with the size of set $H$, the biasing page set, and they criticize previous ideas in~\cite{qd-pagerank}:
``[Richarson and Domingos] suggested that importance scores be precomputed offline for every possible text query, but the enormous number of possibilities makes this approach difficult to scale."

In this paper, we propose a different alternative to the problem of faceted ranking. Instead of computing offline the rankings corresponding to all possible facets, %as proposed in Richarson and Domingos~\cite{qd-pagerank},
our solution requires only the offline computation of a ranking per tag. A faceted ranking is generated by adequately merging the rankings of the corresponding tags. Section \ref{sec:definitions_and_algorithms} deals with different approaches to the merging step.

\section{Construction of a tagged graph}
\label{sec:construction}

In this section we introduce the basic definitions related to tagged graphs and we present the network analysis of two real cases.

\subsection{Basic definitions}
\label{subsec:basic_definitions}

Let $G = (N,E,T)$ be a simple\footnote{Not a \emph{multigraph}.
} directed graph with tags on the edges, a \emph{tagged graph}.
$N$ is the set of vertices $\{ u_1, \ldots, u_n \}$, $E$ is the set of edges  and $T(e)$ is the set of tags $\{t_1, \ldots, t_{k_e} \}$ associated with edge $e$ in $E$.
If $e \notin E$ then $T(e):= \emptyset$.
We shall call a certain set of tags $F\subseteq \bigcup_{e\in E}T(e)$ a \emph{facet}.

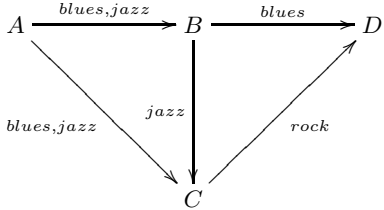
\begin{figure}[t]
\begin{center}
{\small
\begin{tabular}{rlrl}
$M_0$ = &{\em \{(A, song1, \{blues\}) }&$V_0$ =  &{\em \{(A, song2)}  \\
   &{\em (B, song2, \{blues,jazz\})} &   & {\em (B, song4)} \\
   &{\em (C, song3,\{blues\})}       &   & {\em (B, song5)} \\
   &{\em (C, song4,\{jazz\}) }       &   & {\em (A, song3)} \\
   &{\em (D, song5,\{blues\})}       &   & {\em (A, song4)}  \\
   &{\em (D, song6,\{rock\}) \}}     &   & {\em (C, song6) \}} \\
\end{tabular}

\[
\xymatrix{
A \ar[rr]^{blues,jazz} \ar[ddrr]_{blues,jazz}  &  & B \ar[dd]_{jazz} \ar[rr]^{blues} & & D\\
&  & \\
  & &  C\ar[uurr]_{rock} & & \\
}
\]
}
\end{center}
\caption{Example of construction of a tagged graph from a set of contents $M_0$ and a set of recommendations $V_0$.}
\label{fig:example_1}
\end{figure}

Let  $M = \{ (u_1, m_1,T_1), \ldots, (u_r, m_r,T_r) \}$ be a set of tagged contents, where $u_i$ is the user, $m_i$ is the content
and $T_i$ is the preferred set of tags included by
the user\footnote{In the rest of the paper \emph{user} or \emph{vertex} will be used indistinctly to mean a \emph{vertex} in a tagged graph as an abstraction of the webpage where the user publishes his/her content, and \emph{edges} or \emph{links} will be used when referring to edges in a tagged graph build using favorite recommendations.}, and let $V = \{ (c'_1, m'_1), \ldots, (c'_p, m'_p) \}$ the set of favorite recommendations, where $c'_i$ is a recommender user and $m'_i$ is the recommended content\footnote{Each content is considered unique, i.e., different users do not upload the same content.}, then a tagged graph $G = (N,E,T)$ is build, where
\begin{multline*}
N := \{ u_i : \exists i (u_i,m_i,T_i) \in M  \} \cup\\
    \{ c'_j : \exists j (c'_j,m'_j) \in V  \},
\end{multline*}
\[
E := \{ (c'_j,u_k) : (c'_j, m'_j) \in V \land (u_k,m'_j,T_k) \in M  \},
\]
and
\begin{multline*}
T((c'_j, u_k)) := \{ T_k :  (c'_j, m'_j) \in V \land \\ (u_k,m'_j,T_k) \in M \land
 (c'_j, u_k) \in E\} \enspace.
\end{multline*}
We show an example of the application of these definitions in Figure~\ref{fig:example_1}.

Given $G = (N,E,T)$ and a tag $t$ then $G(t) := (N', E', T')$ is a tagged subgraph, where
$E' = \{ e : e \in E \land t \in T(e) \}$,
$N' = \{ a,b : (a,b) \in E' \}$ and $T' = \{t\in T(e'): e' \in E'\}$.

If $G_1 = (N_1,E_1,T_1)$ and $G_2 = (N_2,E_2,T_2)$ are graphs then
$G_1 \cap G_2 := (N', E', T')$ where
$E' = E_1 \cap E_2$,
$N' = \{ a,b : (a,b) \in E' \}$ and $T'(e) := T_1(e)\cap T_2(e)$.
Also
$G_1 \cup G_2 := (N', E', T')$ where
$E' = E_1 \cup E_2$,
$N' = \{ a,b : (a,b) \in E' \}$  and $T'(e) := T_1(e)\cup T_2(e)$.

The conjunction and disjunction graphs can be defined as $G(t_1 \land \ldots \land t_{k-1} \land t_k) := G(t_1 \land \ldots \land t_{k-1}) \cap G(t_k)$
 (see Figure~\ref{fig:example_1_bis_bluesandjazz})
 and $G(t_1 \lor \ldots \lor t_{k-1} \lor t_k) := G(t_1 \lor \ldots \lor t_{k-1}) \cup G(t_k)$
 (see Figure~\ref{fig:example_1_bis_bluesorjazz})
.

The number of edges of a graph $G$ is denoted $|E(G)|$ and the number of vertices in a graph is denoted by $|N(G)|$.

\subsection{Two real cases: YouTube and Flickr}
\label{subsec:examples}

In this section, we present two examples of collaborative tagging systems where content is tagged and recommendations are made.
These systems actually rank content according to the number of visits, recommendations or relevance of the text accompanying the content. However, to our knowledge, no use of graph-based faceted ranking is made.

The taxonomy of tagging systems in Marlow {\em et al.}~\cite{tagging} allows us to classify \emph{YouTube}~\cite{youtube} and \emph{Flickr}~\cite{Flickr} in the following ways:
\begin{itemize}
\item
regarding the tagging rights, both are \emph{self-tagging} systems;
\item
 regarding the aggregation model, they are \emph{set} systems;
\item
 regarding the object-type, they are called \emph{non-textual} systems;
\item
 regarding source of material, they are classified as \emph{user-contributed};
\item
 finally, regarding tagging support, while \emph{YouTube} can be classified as a \emph{suggested} tagging system, \emph{Flickr} must be considered a \emph{blind} tagging system.
\end{itemize}

In our first example the content is multimedia in the form of favorite videos recommended by users. The information was collected from the service \emph{YouTube}~\cite{youtube} using the public API crawling 185852 edges and 51490 vertices in Breadth-First Search (BFS) order starting from the popular user \emph{jcl5m} that had videos included in the top twenty top rated videos during April 2008. From this information and following the the definitions in Section \ref{subsec:basic_definitions}, we constructed a complete tagged graph $G$ and several sample subgraphs such as $G(music \lor funny)$, $G(music)$, $G(funny)$ and $G(music \land funny)$ (other subgraphs present a similar behavior). Table~\ref{tab:video_graphsizes} presents the number of vertices and edges of each of these networks. We must note that mandatory categorical tags such as $Entertainment$, $Sports$ or $Music$, always capitalized, were removed in order to include only tags inserted by users.

\begin{table}[ht]{%\footnotesize
\begin{center}
\begin{tabular}{ l | r | r  }
  \hline
  Graph & vertices & edges \\ \hline \hline
  $G$ & 51,490 & 185,852 \\
  $G(music \lor funny)$ & 18,368 & 26,388 \\
  $G(music)$ & 12,849 & 10,273 \\
  $G(funny)$ & 8,734 & 13,392 \\
  $G(music \land funny)$ & 1,406 & 1,147 \\
  \hline
\end{tabular}
\end{center} }
\caption{Sizes of the video tagged graph and some of its subgraphs.}
\label{tab:video_graphsizes}
\end{table}

In our second example the content are photos and the recommendations are in the form of favorite photos\footnote{Only the first fifty favorites photos of each user were retrieved.}.
The information was collected from the service \emph{Flickr}~\cite{Flickr} using the public API crawling 229709 edges and 35210 vertices in BFS order starting from the popular user \emph{junku-newcleus}. The complete tagged graph $G$ and the sample subgraphs $G(blue \lor flower)$, $G(blue)$, $G(flower)$ and $G(blue \land flower)$ were constructed. The number of vertices and edges of these graphs are shown in Table~\ref{tab:photo_graphsizes}.

\begin{table}[ht] {%\footnotesize
\begin{center}
\begin{tabular}{ l | r | r  }
  \hline
  Graph & vertices & edges \\ \hline \hline
  $G$ & 35,210 & 229,709 \\
  $G(blue \lor flower)$ & 12,921 & 20,105 \\
  $G(blue)$ & 10,241  & 11,703  \\
  $G(flower)$ & 7,032 & 9,566 \\
  $G(blue \land flower)$ & 1,551 & 1,164 \\
  \hline
\end{tabular}
\end{center} }
\caption{Sizes of the photo tagged graph and some of its subgraphs.}
\label{tab:photo_graphsizes}
\end{table}

\begin{figure}[]
\begin{center}
\includegraphics[scale=0.32,angle=-90]{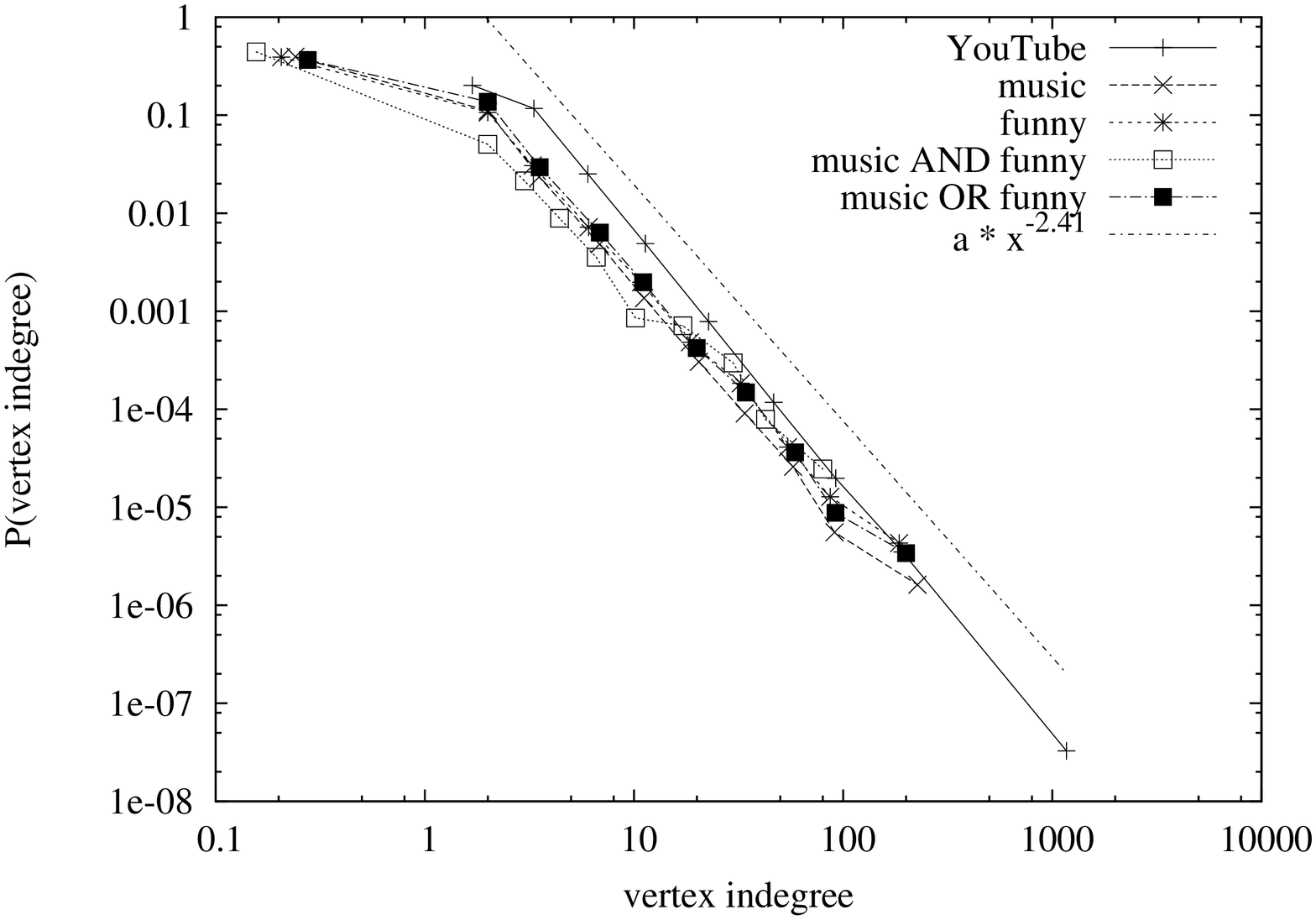}
\includegraphics[scale=0.32,angle=-90]{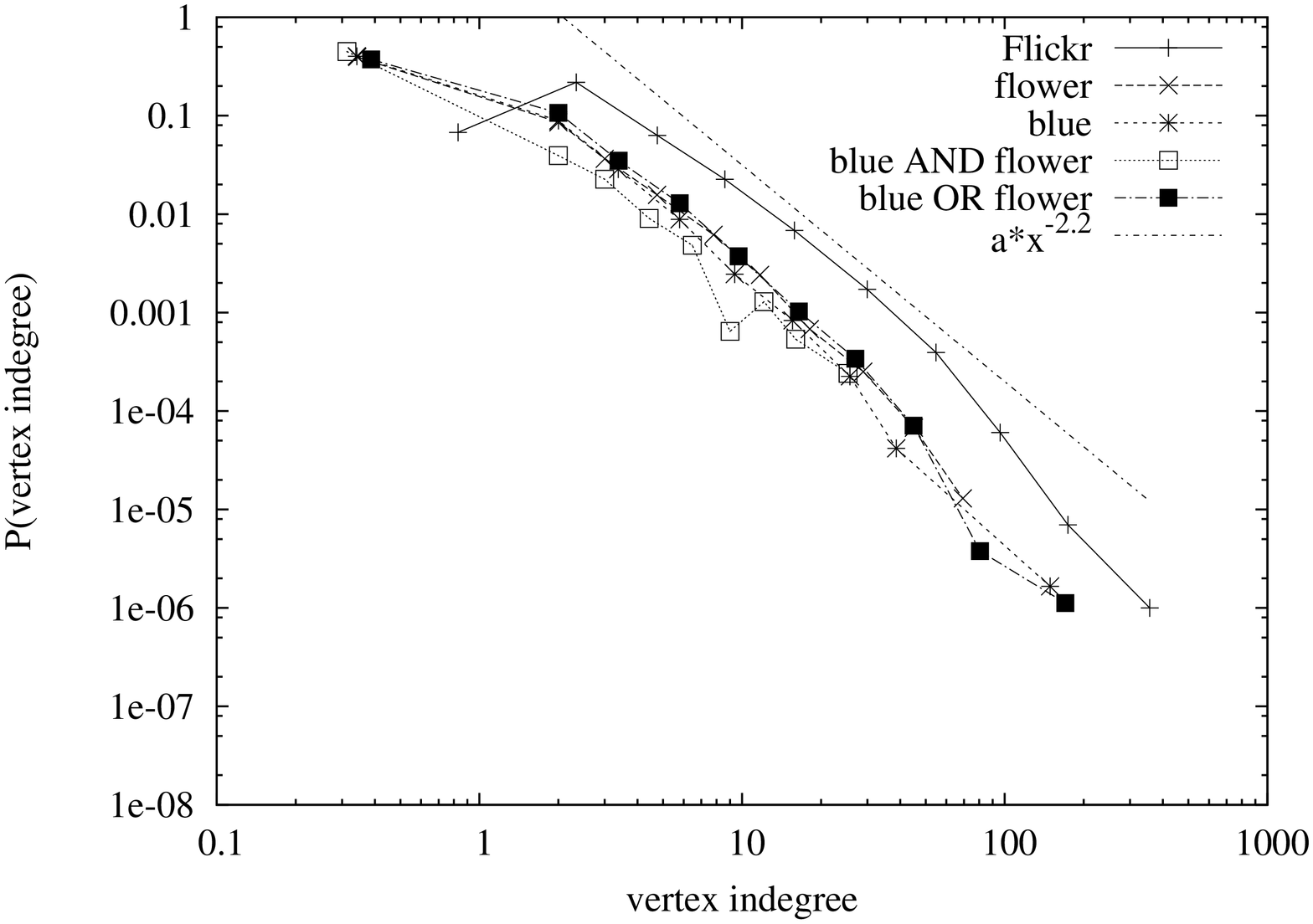}
\caption{Binned indegree distribution}
\label{fig:netinfo_indegree}
\end{center}
\end{figure}
\begin{figure}[]
\begin{center}
\includegraphics[scale=0.32,angle=-90]{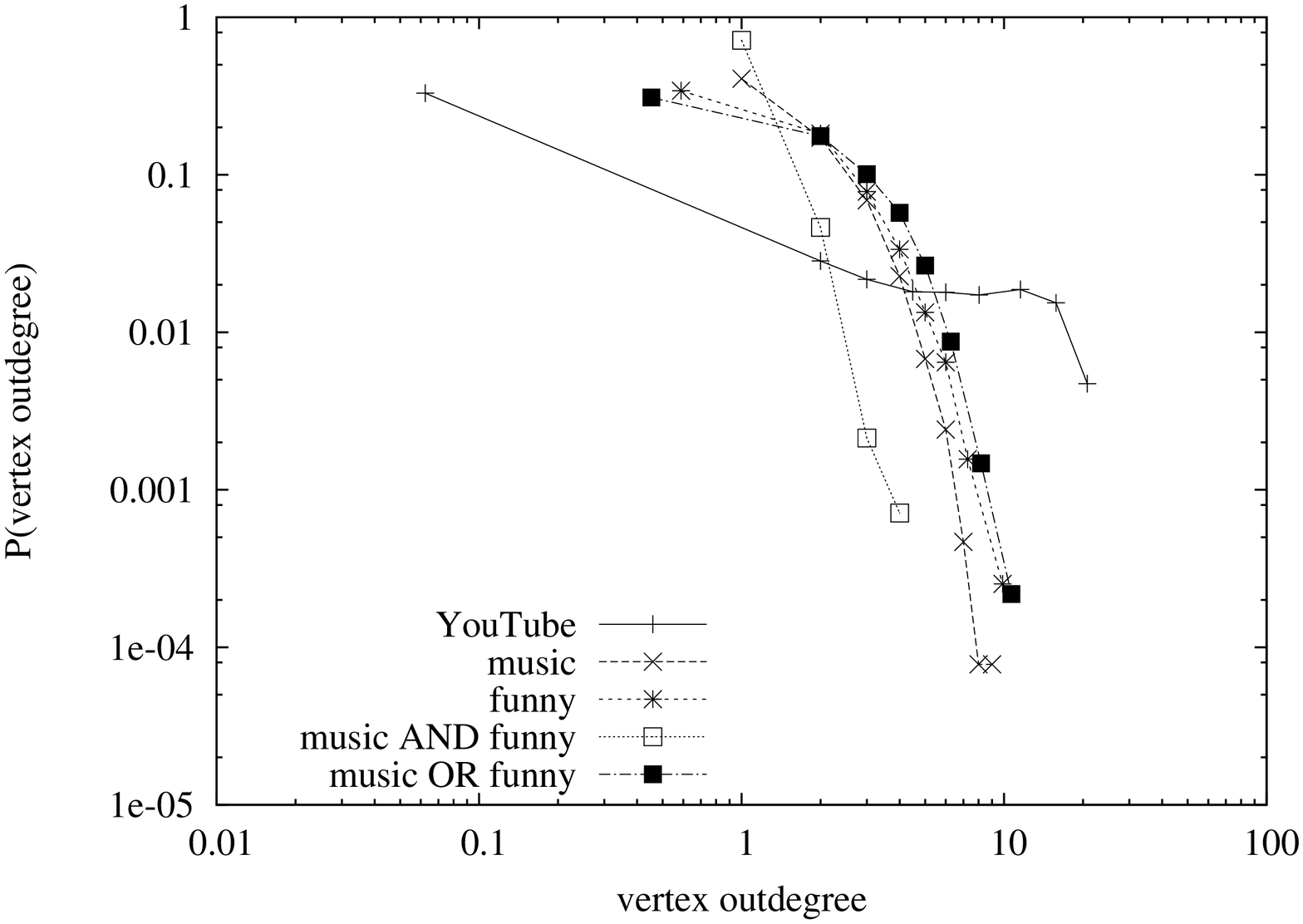}
\includegraphics[scale=0.32,angle=-90]{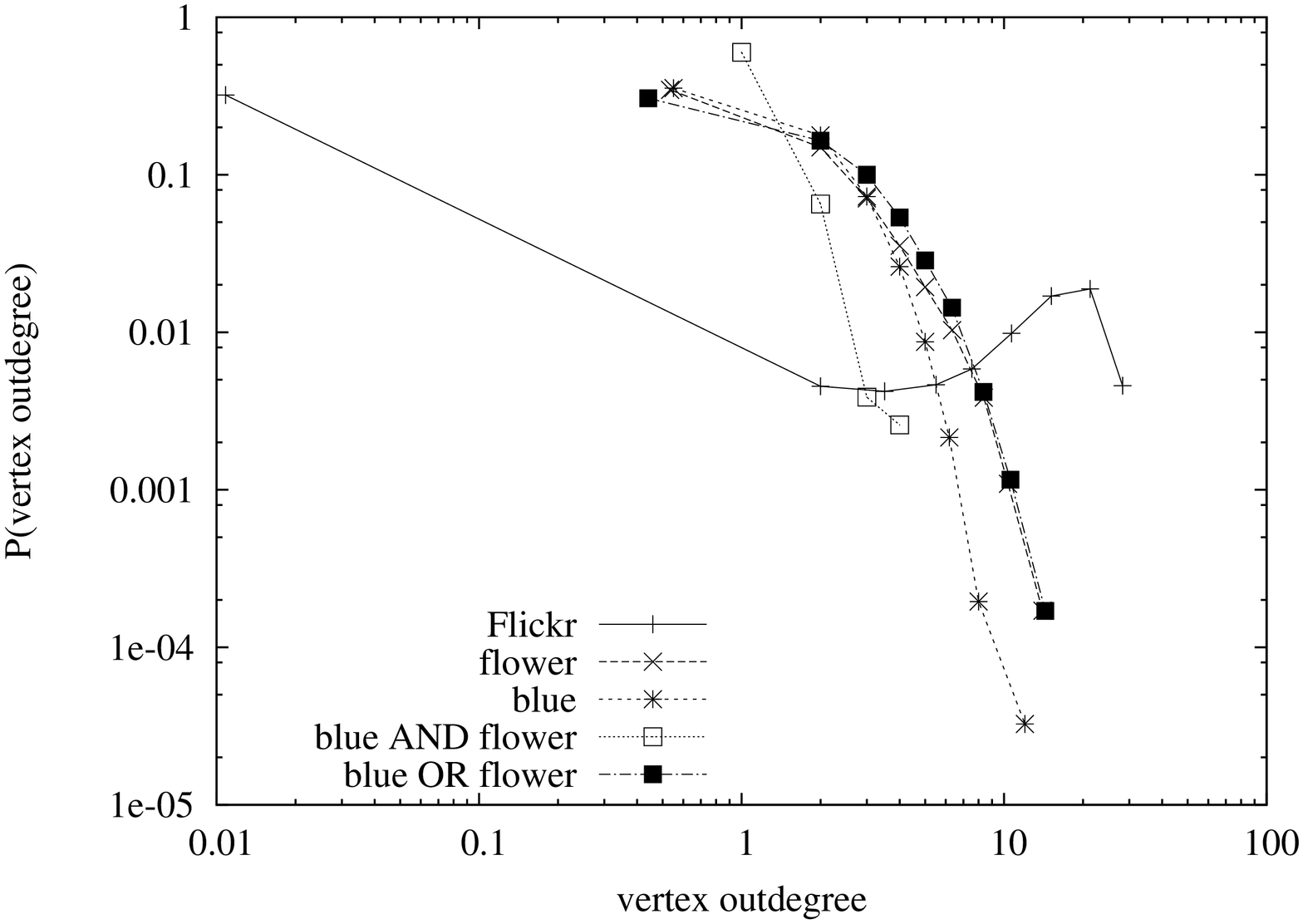}
\caption{Binned outdegree distribution}
\label{fig:netinfo_outdegree}
\end{center}
\end{figure}
\begin{figure}[]
\begin{center}
\includegraphics[scale=0.32,angle=-90]{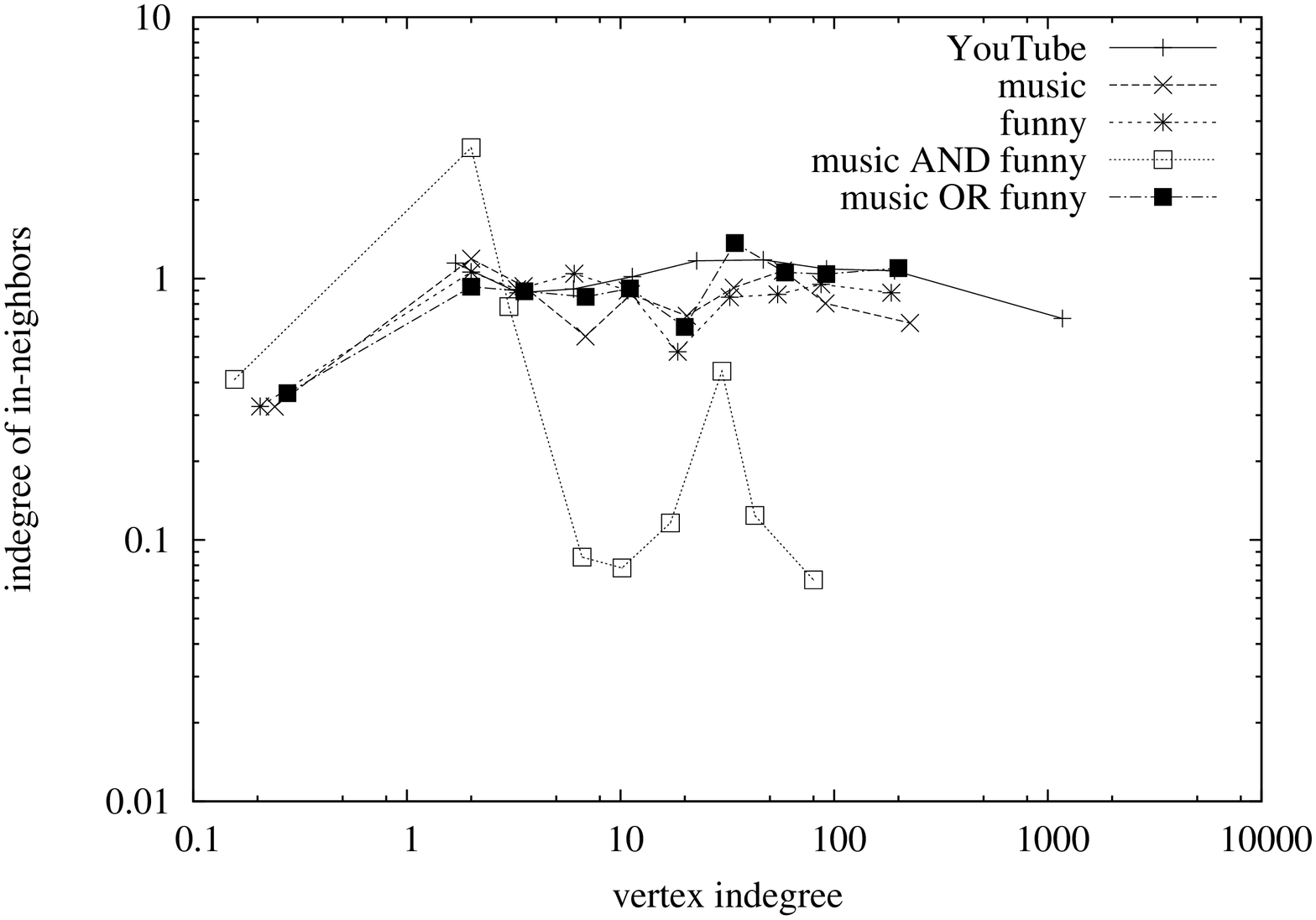} \\
\includegraphics[scale=0.32,angle=-90]{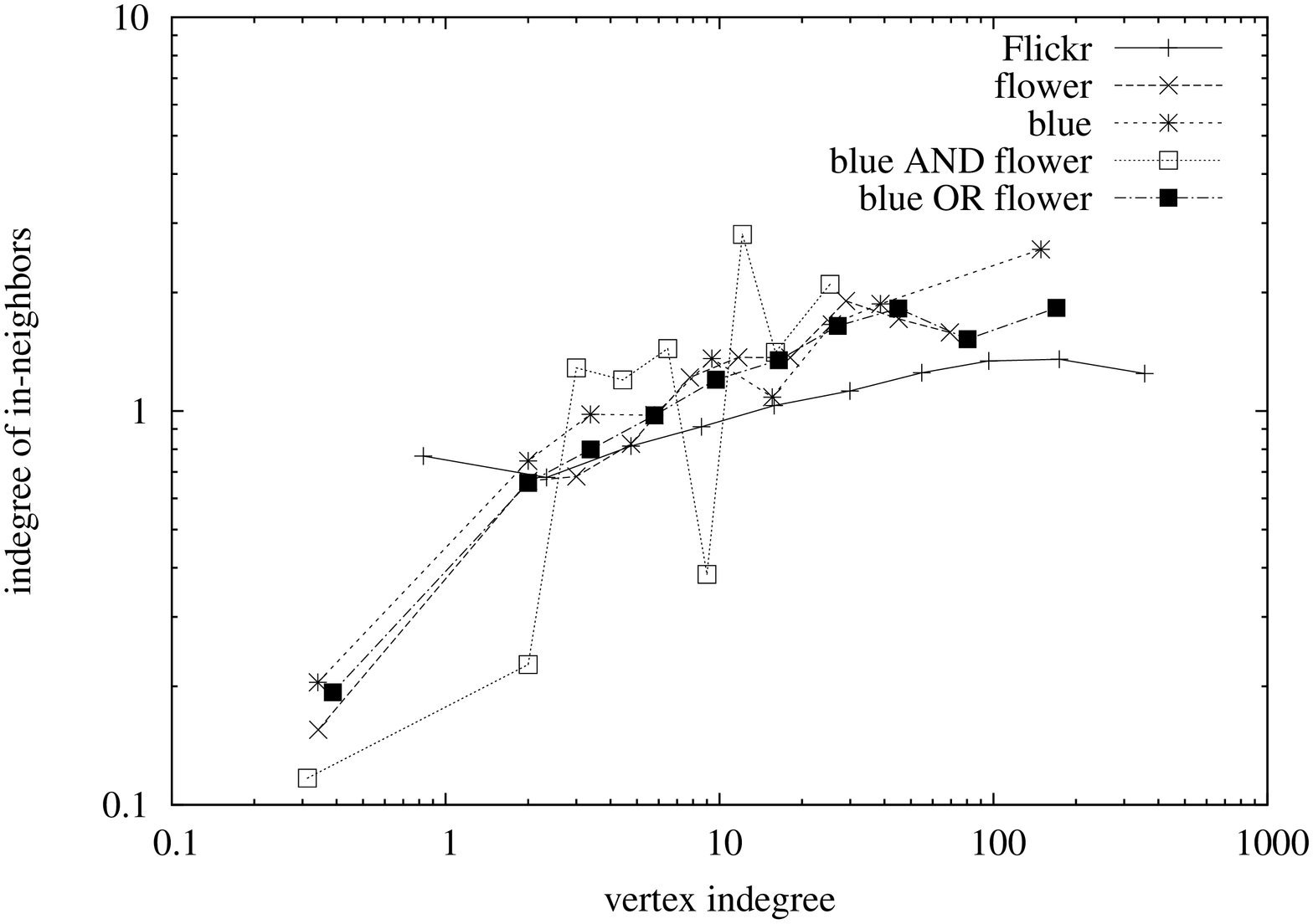}
\caption{Binned correlation of indegree of in-neighbors with indegree}
\label{fig:netinfo_inoutcorrel}
\end{center}
\end{figure}

\begin{figure}
\begin{center}
\includegraphics[scale=0.32,angle=-90]{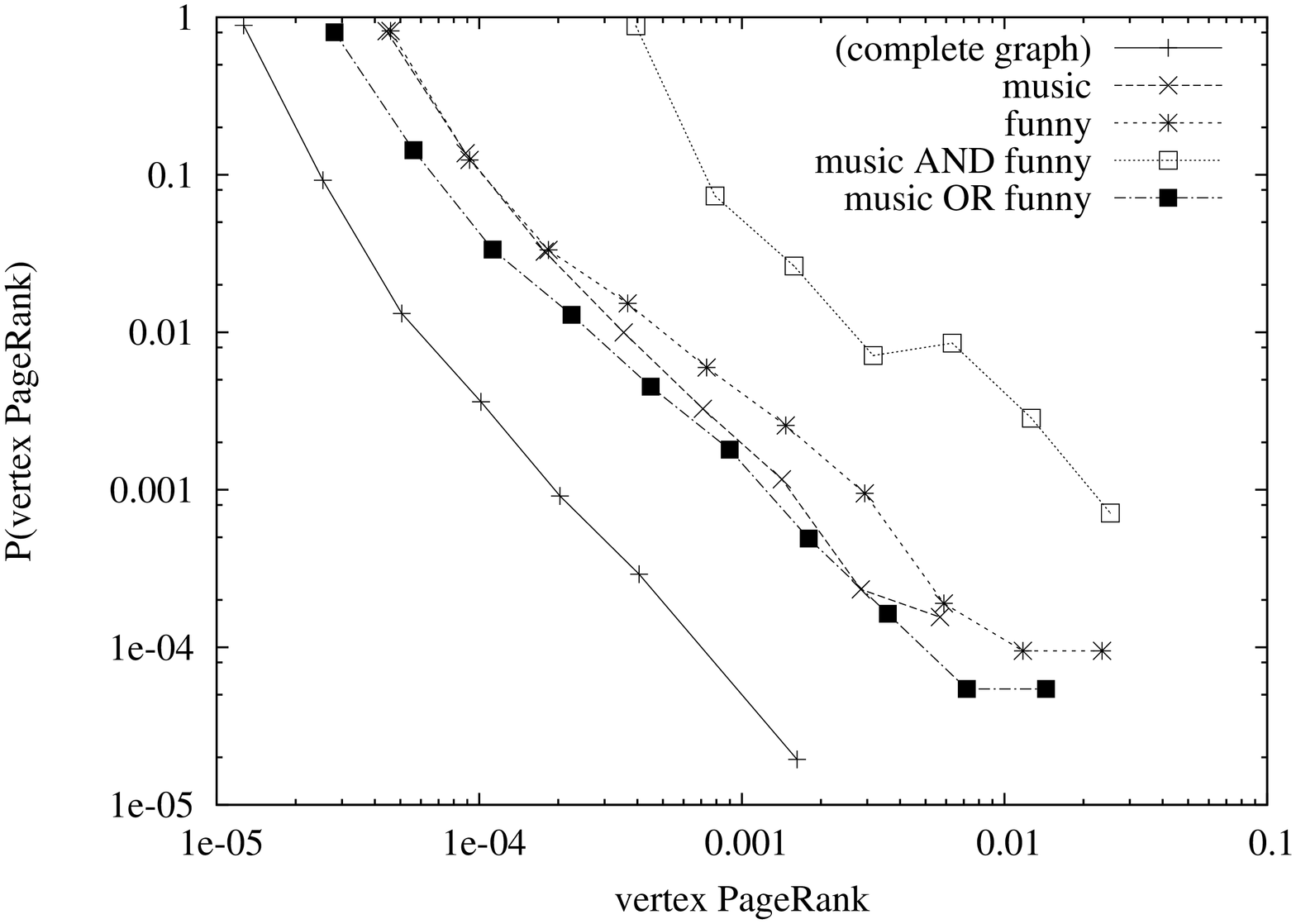}
\includegraphics[scale=0.32,angle=-90]{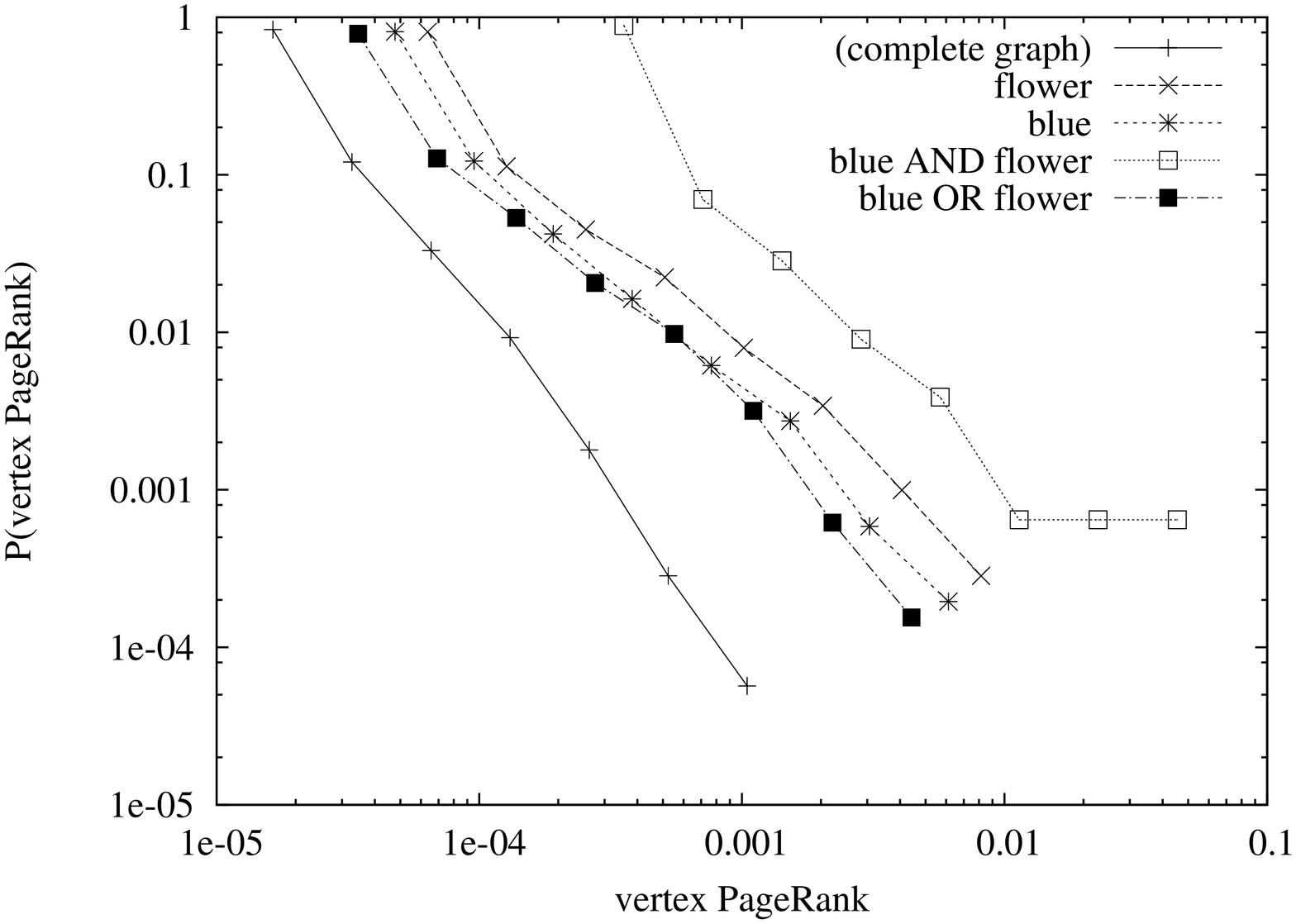}
\caption{Binned Vertex PageRank distribution for YouTube (top) and
Flickr (bottom)}
\label{fig:pagerank_distributions}
\end{center}
\end{figure}

\subsection{Network analysis}
\label{subsec:network-analysis}

Graph analysis was made using the tool Network Workbench~\cite{nwb}, except for the calculation of PageRank.
Figures~\ref{fig:netinfo_indegree}, \ref{fig:netinfo_outdegree} and~\ref{fig:netinfo_inoutcorrel} show vertex indegree distribution, vertex outdegree distribution and correlation of indegree of in-neighbors with indegree of vertices for the {\em YouTube} and {\em Flickr} networks.
All graph-analytical parameters, except those for small subgraphs like $G(music \land funny)$ were binned and plotted in log-log curves.
This is the reason why some degree points appear below zero and one ($x$-axis), because there exist vertices with either indegree or outdegree equal to zero.

Vertex indegree, in both video and photo networks, is characterized by a power-law distribution: $P(k) \approx k^{-\gamma}$, where $2<\gamma<3$ (see Figure~\ref{fig:netinfo_indegree}).
Random variables modelled by this type of heavy-tailed distributions have a finite mean, but infinite second and higher non-central moments.
Furthermore, there is a non-vanishing probability of finding a vertex with an arbitrary high indegree.
Clearly, in any real-world network, the total number of vertices is a natural upper-bound to the greatest possible indegree.
However, experience with Internet related networks shows that the power-law distribution of the indegree does not change significantly as the network grows and, hence, the probability of finding a vertex with an arbitrary degree eventually becomes non-zero (for more details see, e.g., Pastor-Satorras and Vespignani~\cite{psvbook}).

Since recommendation lists are made by individual users, vertex outdegree does not show the same kind of scale-free behavior than vertex indegree. On the contrary, each user recommends only 20 to 30 other users on average (see Figure~\ref{fig:netinfo_outdegree}). Moreover, since vertex outdegree is mostly controlled by {\em human} users, we do not expect its average to change significantly as the network grows.

The correlation of indegree of in-neighbors with vertex indegree (see Figure~\ref{fig:netinfo_inoutcorrel}) indicates the existence of assortative (positive slope) or disassortative behavior (negative slope).
Assortativeness is commonly observed in social networks, where people with many connections relates to people which is also well-connected. Disassortativeness is more common in other kinds of networks, such as information, technological and biological networks (see, e.g., Newman~\cite{newman}).
In the favorite videos network there is no clear correlation (small or no slope), but the photo network there is a slight assortativeness indicating a biased preference of vertices with high indegree for vertices with high indegree (see Figure~\ref{fig:netinfo_inoutcorrel}).

We also computed the PageRank of the sample graphs, removing dangling vertices with indegree 1 and out degree 0, because most of them correspond to vertices which have not been expanded by the crawler (BFS), having the lowest PageRank (a similar approach is taken in~\cite{pagerank_original}).
Figure~\ref{fig:pagerank_distributions} shows that PageRank distributions are also scale-free, i.e., they can be approximated by power law distributions.
Note that the power law exponents are very similar for the complete tagged graph and subgraphs, on each network.

\section{Faceted Ranking on Tagged Graphs}
\label{sec:definitions_and_algorithms}

Given a set $M$ of tagged content, a set $V$ of favorite recommendations and a tag set or facet $F$, the  \emph{faceted ranking problem} consists in finding the ranking of users according to facet $F$.

In this section we present six different approaches to the faceted ranking problem using tagged graphs.
The first two algorithms ($E$-intersection and $E$-union/$N$-intersection in Sections~\ref{sec:e-intersection} and~\ref{sec:e-union-n-intersection} respectively) are not scalable for online queries because their computation requires the extraction of a subgraph which might be very large in a large network\footnote{We have observed that as the network grows the relative frequency of tags usage converges. Similar behavior was observed for particular resources by~\cite{folksonomies}.} and the calculation of the corresponding PageRank vector. Moreover, the offline computation of those rankings for each possible facet $F\subseteq \bigcup_{e\in E}T(e)$ is also unfeasible because the large number of such facets. However, they serve as a basis of comparison for the other four online algorithms because they are a good approximation to the desired result.

We should note that the focus of this paper is on  \emph{conjunction-based} queries in which \emph{all words} must be matched, as opposed to \emph{disjunction-based} ones where  the match of \emph{any word} is sufficient.
Conjunction-based queries are the most common type of boolean queries~\cite{ir_book}.

Before presenting faceted ranking algorithms, we need some preliminary definitions related to vertex centrality which are given in the following section.

\subsection{Vertex Centrality}

Given a graph $G = (N, E)$, $C(G): N \rightarrow \mathbb{R}$ is a \emph{vertex centrality function} and $R(C(G)) : N \rightarrow \mathbb{N}$ is a \emph{vertex ranking function} associates a complete order such that to the highest centrality vertex of $C(G)$  corresponds the number one, the second highest has number two and so on.
PageRank $\mathcal{C}(G)$ is a vertex centrality function associating probabilities according to a random surfer traversing the graph $G$~\cite{pagerank}.
The vertex ranking function $\mathcal{R}(G) := R(\mathcal{C}(G))$ will be our default ranking for graphs.

 \begin{figure}[ht]
 {
 \begin{center}
 {\small
 {
 \subfigure[]{
 $$
 \xymatrix{
 A \ar[rr]^{blues,jazz} \ar[ddrr]_{blues,jazz} & & B\ar[rr]^{blues} & & D\\
 & &  & & \\
 &  & C & &\\
 }
 $$
 \label{fig:example_1_bis_blues}
 }

 \subfigure[]{
 $$
 \xymatrix{
 A \ar[rr]^{blues,jazz} \ar[ddrr]_{blues,jazz} & &  B\ar[dd]^{jazz}\\
 & & \\
   && C   \\
 }
 \label{fig:example_1_bis_jazz}
 $$
 }
 \subfigure[]{
 $$
 \xymatrix{
 A \ar[rr]^{blues,jazz} \ar[ddrr]_{blues,jazz} & & B \\
 && \\
 & & C  \\
 }
 $$
 \label{fig:example_1_bis_bluesandjazz}
 }

 \subfigure[]{
 $$
 \xymatrix{
 A \ar[rr]^{blues,jazz} \ar[ddrr]_{blues,jazz} & & B \ar[dd]^{jazz} \ar[rr]^{blues} & & D\\
  & &  & &\\
  & & C & &\\
 }
 $$
 \label{fig:example_1_bis_bluesorjazz}
 }
 }
 }
 \end{center}
 \caption{Example subgraphs of a tagged graph: \subref{fig:example_1_bis_blues} $G(blues)$; \subref{fig:example_1_bis_jazz}  $G(jazz)$; \subref{fig:example_1_bis_bluesandjazz} $G(blues \land jazz)$; \subref{fig:example_1_bis_bluesorjazz}} $G(blues \lor jazz)$.
 \label{fig:example_1_bis}
 }
\end{figure}
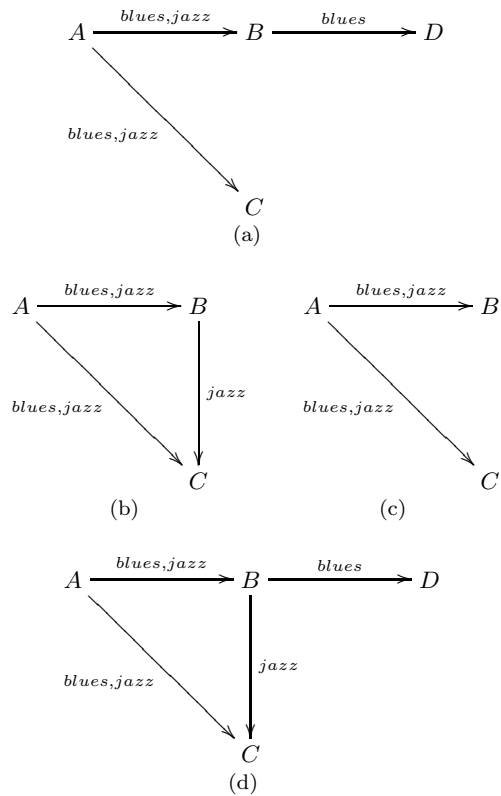

\subsection{$E$-intersection}
\label{sec:e-intersection}

Given a set of tags, a ranking may be calculated by computing the centrality measure of the subgraph corresponding to the recommendation edges which include \emph{all} the tags.
This approach, called $E$-intersection, cannot be implemented for online queries, as explained above, but serves as a reasonable standard of comparison because we use the exact information available for the PageRank in a conjunctive query.

The \emph{$E$-intersection ranking} for tagged graph $G$ according to facet $F = \{ t_1, \ldots, t_k \}$ is
\[
\mathcal{R}(G(t_1 \land \ldots \land t_k)).
\]
As an example see Figure~\ref{fig:example_1_bis_bluesandjazz}.
Assuming a previously built inverted index
(Christopher~\cite{ir_book}, Chapter 1) for the tagged graph mapping
tags into sets of edges, the complexity can be decomposed on the
retrieval time for each subgraph, which takes proportional to
$\sum_{i=1}^k{|E(G(t_i))|}$, and the time of PageRank and sort
algorithms, taking $\mathcal{O}(m \log m)$, where $m = |E(G(t_1
\land \ldots \land t_k))|$. Then, the total time complexity for this
algorithm is $\mathcal{O}\left(k\times|E(G(t_i))|_{\max}+ m\log
m\right)$, where $|E(G(t_i))|_{\max}$ is computed for the largest
subgraph $G(t_i)$.

\subsection{$E$-union/$N$-intersection}
\label{sec:e-union-n-intersection}

Consider the example given in Figure~\ref{fig:example_1} under the query $blues\land rock$. According to the $E$-intersection algorithm, there is no node in the network satisfying the query.
However, it may seem reasonable to return node $D$ as a response to such search. In order to take into account this case, we devised another algorithm called $E$-union/$N$-intersection.
In this case, the union of all edge recommendations per tag is used when computing the PageRank, but only those vertices involved in recommendations for all tags are kept.
The latter filtering is included because we want vertices recommended for each of the tags in the facet.

The \emph{$E$-union/$N$-intersection ranking} for vertex $n$ in a tagged graph $G$ according to facet $F = \{ t_1, \ldots, t_k \}$ is
\[
R(\mathcal{C}(G(t_1 \lor \ldots \lor t_k))) (n),
\]
where $\mathcal{C}$ is restricted to vertices in vertex intersection $N(G(t_1)) \cap \ldots \cap N(G( t_k))$, the other vertices having centrality $0$.
Note that, in general, there are more vertices in $N(G(t_1)) \cap \ldots \cap N(G( t_k))$  than  in $N(G(t_1) \cap \ldots \cap G( t_k))$.

The time complexity of this algorithm is proportional to $\sum_{i=1}^k{|E(G(t_i))|} + m \log m$, where $m = |E(G(t_1 \lor \ldots \lor t_k))|$. Then, the total time complexity for this algorithm is $\mathcal{O}\left(k \times |E(G(t_i))|_{\max}+ m \log m\right)$, where $|E(G(t_i))|_{\max}$ is computed for the largest subgraph $G(t_i)$.

\subsection{Single ranking}

A simple online faceted ranking consists of a monolithic ranking, without considering the facet, which is then filtered to exclude those vertices that are not related to all tags in the facet.
That is, one ranks by the monolithic global rank of the complete tagged graph and the only vertices remaining for facet $\{ t_1, \ldots, t_k \}$ are the ones in
\[
N\big(G(t_1)\big) \cap \ldots \cap N\big(G( t_k)\big).
\]

Assuming a precomputed inverted index, mapping tags into nodes, the complexity of this algorithm is $\mathcal{O}(k\times |N(G(t_i))|_{\max})$, where $k$ is the number of different tags in the facet, and $|N(G(t_i))|_{\max}$ is computed for the biggest subgraph $G(t_i)$.
It is also possible to retrieve a (small) constant number of \emph{top} elements to intersect, yielding a time complexity of $\mathcal{O}(k)$.

\subsection{$PR$-product}
\label{subsec:pr-product}

In order to approximate efficiently the edge intersection we can precompute individual rankings for each tag and then combine them by element-wise multiplication.
This approximation is inspired on the probability product of independent events.

If $G_1$ and $G_2$ are subgraphs of graph $G$ we define \emph{PageRank ranking product} as
\[
\mathcal{R}(G_1) \cdot \mathcal{R}(G_2) := R(\mathcal{C}(G_1) \cdot \mathcal{C}(G_2)),
\]
where $(\mathcal{C}(G_1) \cdot \mathcal{C}(G_2))(n) := \mathcal{C}(G_1)(n) \cdot \mathcal{C}(G_2)(n)$ (real product).
The \emph{$PR$-product ranking} for tagged graph $G$ according to facet $F = \{ t_1, \ldots, t_k \}$ is
\[
\prod_{i=0}^k \mathcal{R}(G(t_i)).
\]

Assuming the individual rankings for all tags have been computed, the complexity of this algorithm is $\mathcal{O}(k\times |N(G(t_i))|_{\max})$.
Here, it is also possible to reduce the time complexity to $\mathcal{O}(k)$ taking a (small) constant number of \emph{top} elements to make the product.

\subsection{$R$-sum}
\label{subsec:r-sum}

Consider a recommendation graph $G$ larger than that in Figure~\ref{fig:example_1} and the query $blues\land jazz$.
Assume that the PageRank of the top three nodes in the rankings corresponding the subgraphs $G(blues)$ and $G(jazz)$ are as given in Table~\ref{tab:example_pr-prod_vs_r-sum}.
Ignoring other nodes, the ranking given by the $PR$-product rule is $a$, $b$ and $c$.
However, it may be argued that node $b$ shows a better equilibrium of PageRank values than node $a$.
Intuitively, one may feel inclined to rank $b$ over $a$ given the values in the table.
In order to follow this intuition, we devised the $R$-sum algorithm which is also intended to avoid \emph{topic drift} inside the queried facet, that is, any tag prevailing over the others.

The \emph{$R$-sum ranking} for a tagged graph $G$ according to facet $F = \{ t_1, \ldots, t_k \}$ is
\[
\sum_{i=0}^k \mathcal{R}(G(t_i)),
\]
where we define \emph{PageRank ranking sum} as
\[
\mathcal{R}(G_1) + \mathcal{R}(G_2) := R(-(\mathcal{R}(G_1) + \mathcal{R}(G_2))).
\]
Notice that in this sum we are using as \emph{centrality} the sum of ranking positions in a reverse order, and according to the $R$-sum algorithm, the ranking of nodes in the example of Table~\ref{tab:example_pr-prod_vs_r-sum} is $b$, $a$ and $c$.

The complexity of this algorithm is similar to that of $PR$-product.

\begin{table}[ht]
\begin{center}
{\small%\footnotesize
\begin{tabular}{c|c|c|c|c} \hline
Node & $\mathcal{C}(G(blues))$ & $\mathcal{C}(G(jazz))$ & $PR$-pr. & $R$-sum \\
\hline \hline
$a$ & 0.75 & 0.04 & 0.03 & 4\\
$b$ & 0.1 & 0.1 & 0.01 & 3 \\
$c$ & 0.01 & 0.05 & 0.005 & 6 \\
\hline
\end{tabular}
}
\end{center}

\caption{Comparison of $PR$-product and $R$-sum in an example.}
\label{tab:example_pr-prod_vs_r-sum}
\end{table}

\subsection{$\tau$-$N$-intersection}
\label{subsec:tau-n-intersection}

In this case, edge intersection is computed involving only vertices (and associated edges) that are on the top $w$ positions of the individual rankings. The \emph{$\tau$-$N$-intersection ranking} for tagged graph $G$ according to facet $F = \{ t_1, \ldots, t_k \}$ is
\[
\mathcal{R}\left(\bigcap_{i=0}^k G(\tau(\mathcal{R}(G(t_i)),w) )(t_i) \right),
\]
where $\tau(\mathcal{R}(G), w)$ is the set of vertices including the top $w$ vertices ranked using PageRank and $G(\{a,b,\ldots\})$ is the maximal subgraph of $G$ including vertices $\{a,b,\ldots\}$ and edges connecting them.
In other words, this algorithm has the following steps:
(i) for each $t_i$, the subgraph $G(t_i)$ is constructed;
(ii) a ranking of users is computed on the basis of the PageRank of $G(t_i)$;
(iii) the $w$ winners of each $t_i$-associated ranking are extracted;
(iv) given a facet $F = \{t_1, \cdots,t_k\}$, a new subgraph including only the winners for tag $t_i$ is constructed;
(v) a facet-associated ranking is constructed based on the new graph.
Steps (i)-(iii) are computed offline. In this presentation, we have fixed the number of top items selected at five hundred ($w=500$).

Assuming the individual rankings for $k$ tags has been computed, the complexity of this algorithm is $\mathcal{O}(k)$.

\subsection{Scalability Analysis}
\label{subsec:scalability}

As noticed by Langville and Meyer~\cite{pagerank}, the number of iterations of PageRank is fixed when both the tolerated error and other parameters are fixed, yielding one hundred for $\epsilon=10^{-6}$ (see Section~\ref{sec:related}).
As each iteration consists of the sparse adjacency matrix multiplication, the time complexity of PageRank is linear on the number of edges of the graph.
In our case, given a tagged graph $G_0 = (N_0,E_0,T_0)$, for each tag there is a corresponding subgraph with a known size. Then the total temporal and spatial complexity of the faceted PageRank for all individual tags is

$$
\sum_{t \in T_0'} | E(G_0(t)) | = \sum_{e \in E_0}|T_0(e)|,
$$
where $T_0' := \bigcup_{e \in E_0}T_0(e)$, the complete set of tags.

\begin{figure}[ht]
\begin{center}
\includegraphics[scale=0.27,angle=-90]{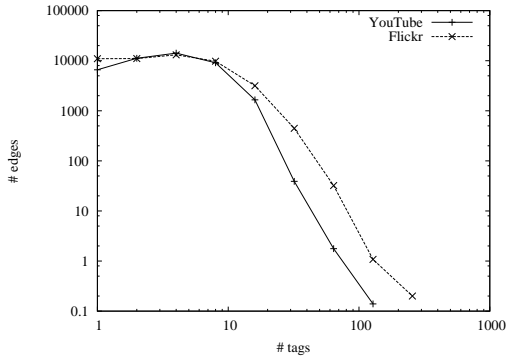}
\end{center}
\caption{The distribution of number of tags per edge.}
\label{fig:tags_hist}
\end{figure}

Therefore, if the average number of tags per edge is constant or grows very slowly as the graph grows, then the algorithms in Sections \ref{subsec:pr-product}, \ref{subsec:r-sum} and \ref{subsec:tau-n-intersection} are scalable, linear on the number of edges of the complete tagged graph.
This can be verified empirically on Figure~\ref{fig:tags_hist}, showing that distribution of tags per edges falls quickly, having a mean of $9.26$ tags per edge for the YouTube tagged graph and $13.37$ for the Flickr tagged graph.
These are \emph{not} heavy-tailed distributions and, since tags are manually added to each uploaded content, we do not expect the average number of tags  per recommendation to increase significantly with network growth.

In our experiments the computation of all the faceted singleton tag rankings ($104,927$ tags) for the video network sample took $211.4$ times more time than the single ranking for the complete tagged graph. Meanwhile the photo network sample ($283,093$ tags) took $1744.9$ times more time.

Our merging algorithms work in real-time because they use only the top $w$ results, where $w$ is a small fixed number like $500$ or $1000$.
Choosing an appropriate $w$ for an application\footnote{How to choose a good $w$ is beyond the scope of this paper.} will enable it to store only the $w$ top elements of each single-tag facet.

\section{Experimental results}
\label{sec:experiments}

In this section, we compare the behavior of the algorithms presented in Section~\ref{sec:definitions_and_algorithms}.
As a basis of comparison we use two algorithms whose online computation is unfeasible, but which are intuitively reasonable: $E$-intersection (Section~\ref{sec:e-intersection}) and $E$-union/$N$-intersection (Section~\ref{sec:e-union-n-intersection}).
In order to quantify the ``distance'' between the results given by two different algorithms, we use two ranking similarity measures, \emph{OSim}~\cite{topic} and \emph{KSim}~\cite{ksim,topic}.
The first measure, $OSim(r_1,r_2)$ indicates the degree of overlap between the top $n$ elements of rankings $r_1$ and $r_2$.
We define the overlap of two sets $A$ and $B$ (each of size $n$) to be ${|A \cap B|}/{n}$.
The second measure, $KSim(r_1,r_2) = {| (u,v) : r_1',r_2' \textrm{ \tt \footnotesize same order } \forall (u,v),u\neq v|}\;\big/\;{|U|(|U|-1)}$ where $U$ in the union of all elements in rankings $r_1$ and $r_2$, $r_1'$ is $r_1$ extended with $U-r_2$ and $r_2$ is extended analogously to obtain $r_2'$.
This measure is a variant of Kendall's distance that considers the relative orderings, i.e., counts how many inversions are in a determined top set.
In both cases, values closer to 0 mean that the results are not similar and closer to 1 mean the opposite.

\subsection{Favorite videos network}

Samples include all facets of tag pairs $\{t_j, t_k\}$ extracted from the $99$ most used tags of the network\footnote{Some tags  like \emph{you}, \emph{video} or \emph{youtube} which give no information were removed from the experiment.}.
That is, $4851$ tag pairs compared with their similarities averaged.
For each tag pair the proposed merging algorithms ($PR$-product, $R$-sum and $\tau$-$N$-intersection) were compared with the reference algorithms ($E$-intersection and $E$-union/$N$-intersection) using $OSim$ and $KSim$ to measure the rankings' similarity.
Some of the tags are:
{\em music, funny, comedy, live, guitar, rock, super, dance, animation, parody, song, mario, game, new, tv, pop, john, love, world.}

Table~\ref{tab:videos_comparison} presents a summary of the comparisons for the favorite videos network, where we display averaged similarities for different top sizes of ranked users.
 Figures~\ref{fig:videos_gray_e-intersec} and~\ref{fig:videos_gray_e-union} also show a more detailed summary of results for the $OSim$ metric (because it discriminates different situations better than $KSim$).
The $x$-axis corresponds to the number of vertices resulting from the basis of comparison algorithm ($E$-intersection or $E$-union/$N$-intersection) and the $y$-axis to the top number $n$ of vertices used to compute the similarities.
The similarity results (between 0 and 1) falling in each of the log-log ranges were averaged.
Observe that darker tones correspond to values closer to 1, i.e., more similar results. White spaces correspond to cases for which there are no data, e.g., whenever the $y$ coordinate is greater than intersection size.

\subsection{Favorite photos network}

Experiments with {\em Flickr} were similar, top $99$ tags paired to form $4851$ tag pairs.
A small sample of the top $99$ tags is:
{\em bw, portrait, nature, bravo, sky, blue, water, soe, flower, light, clouds, sunset, red, film, macro, white, landscape, green, girl, blackandwhite}.

\begin{figure}[H!]
\begin{center}
\includegraphics[height=40mm,angle=0]{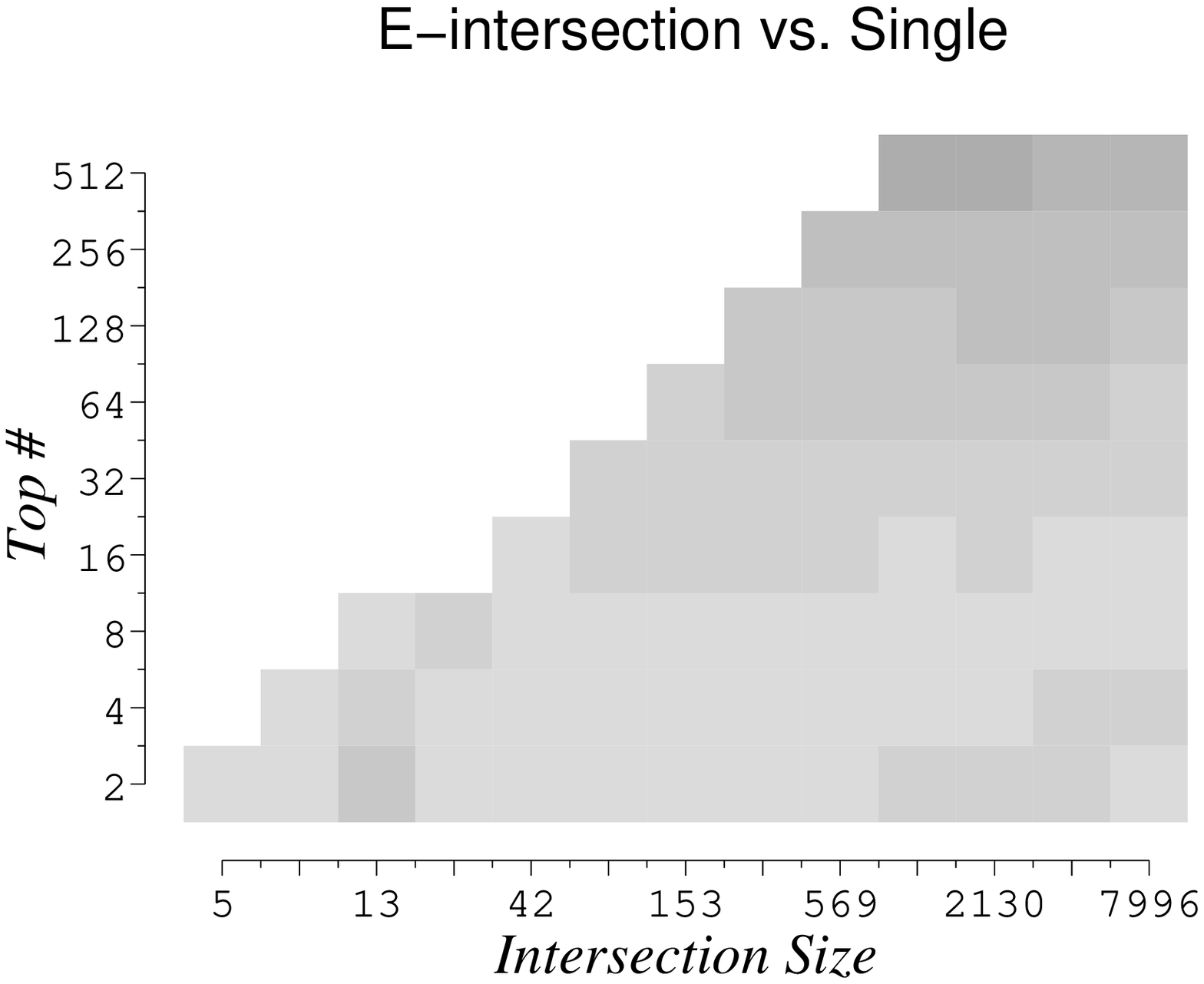}
\includegraphics[height=40mm,angle=0]{bar}
\includegraphics[height=40mm,angle=0]{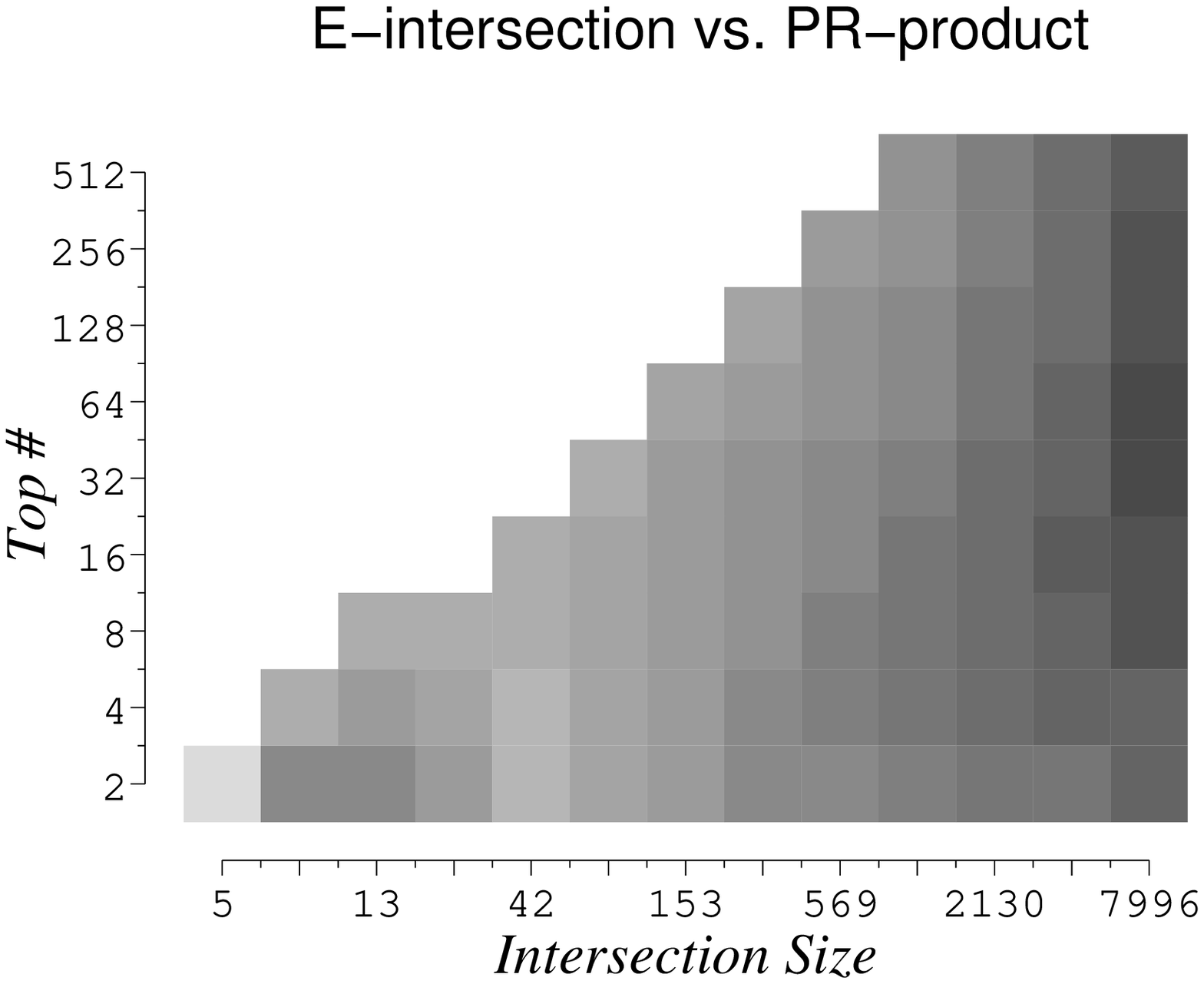}

\includegraphics[height=40mm,angle=0]{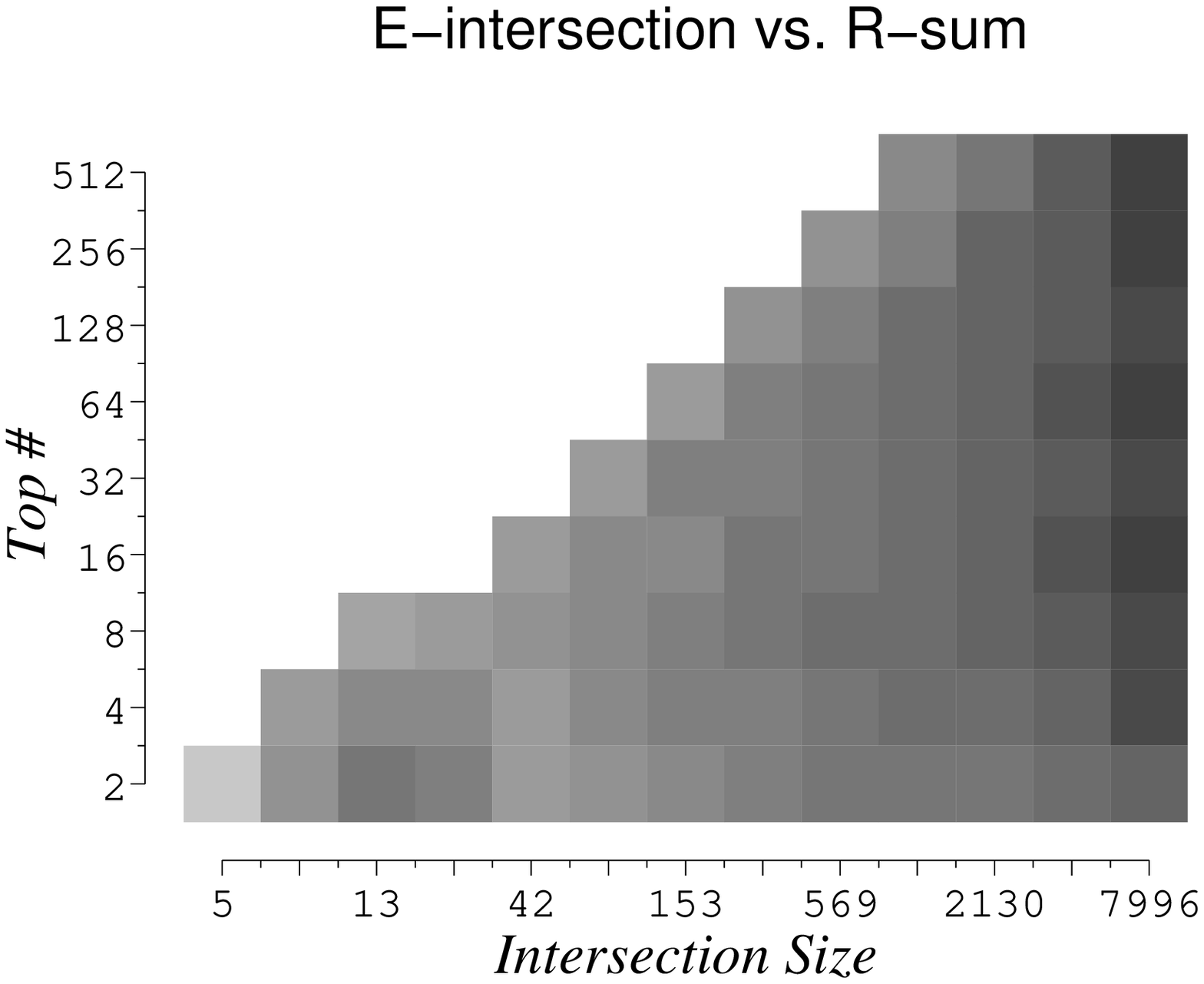}
\includegraphics[height=40mm,angle=0]{bar}
\includegraphics[height=40mm,angle=0]{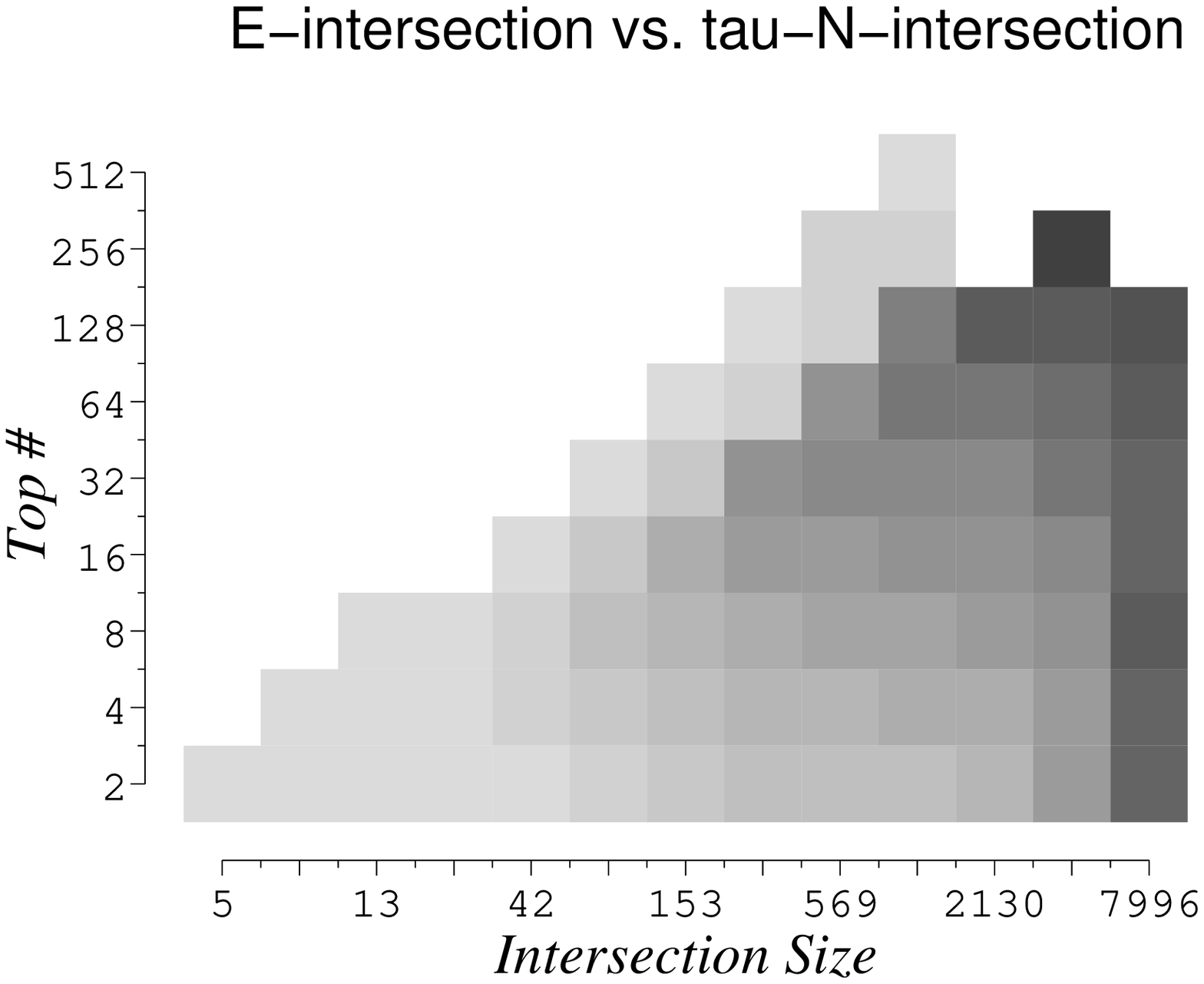}
\end{center}
\caption{Photos network: Average similarity to $E$-intersection}
\label{fig:photos_gray_e-intersec}
\end{figure}

\begin{figure}[H!]
\begin{center}
\includegraphics[height=40mm,angle=0]{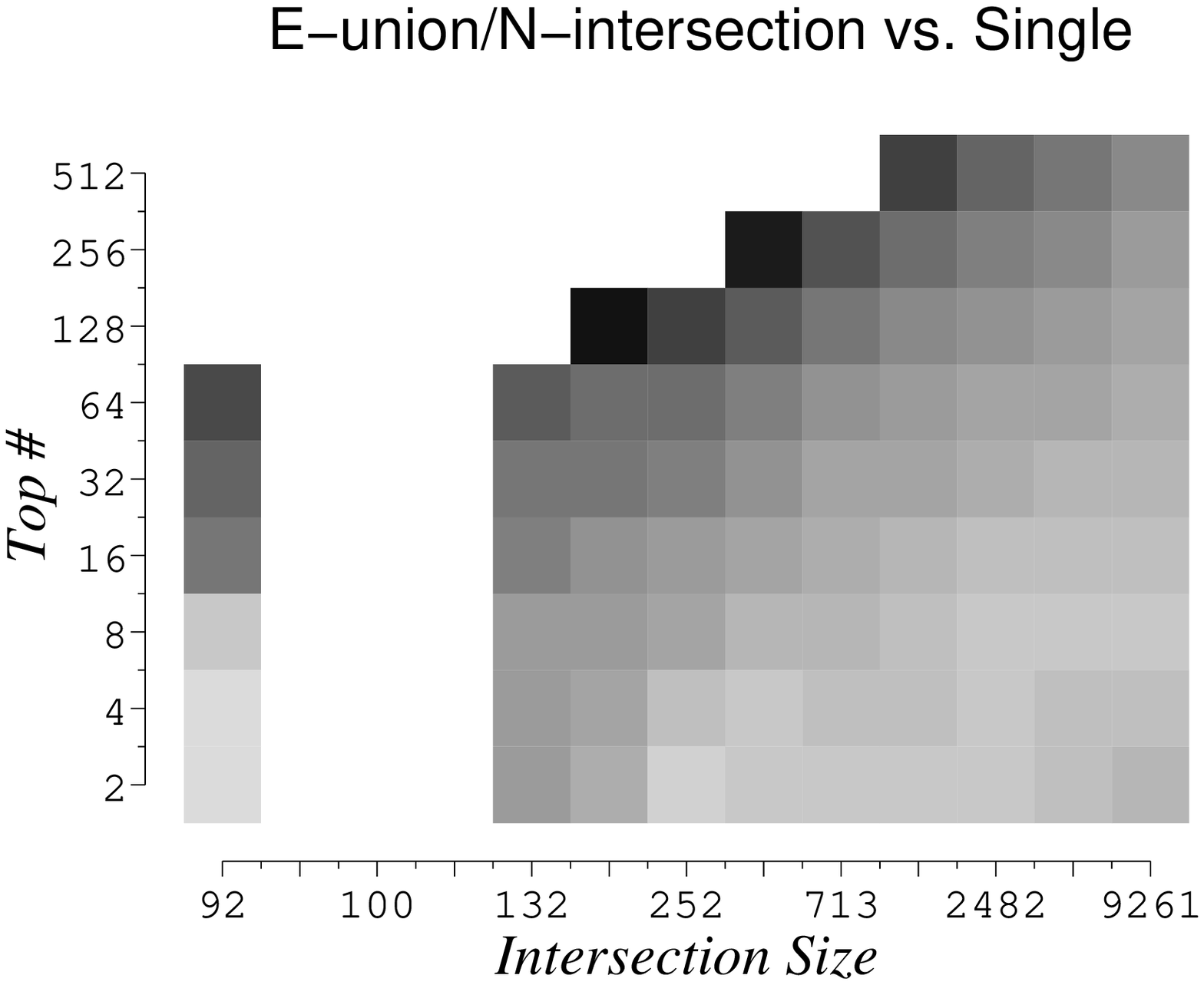}
\includegraphics[height=40mm,angle=0]{bar}
\includegraphics[height=40mm,angle=0]{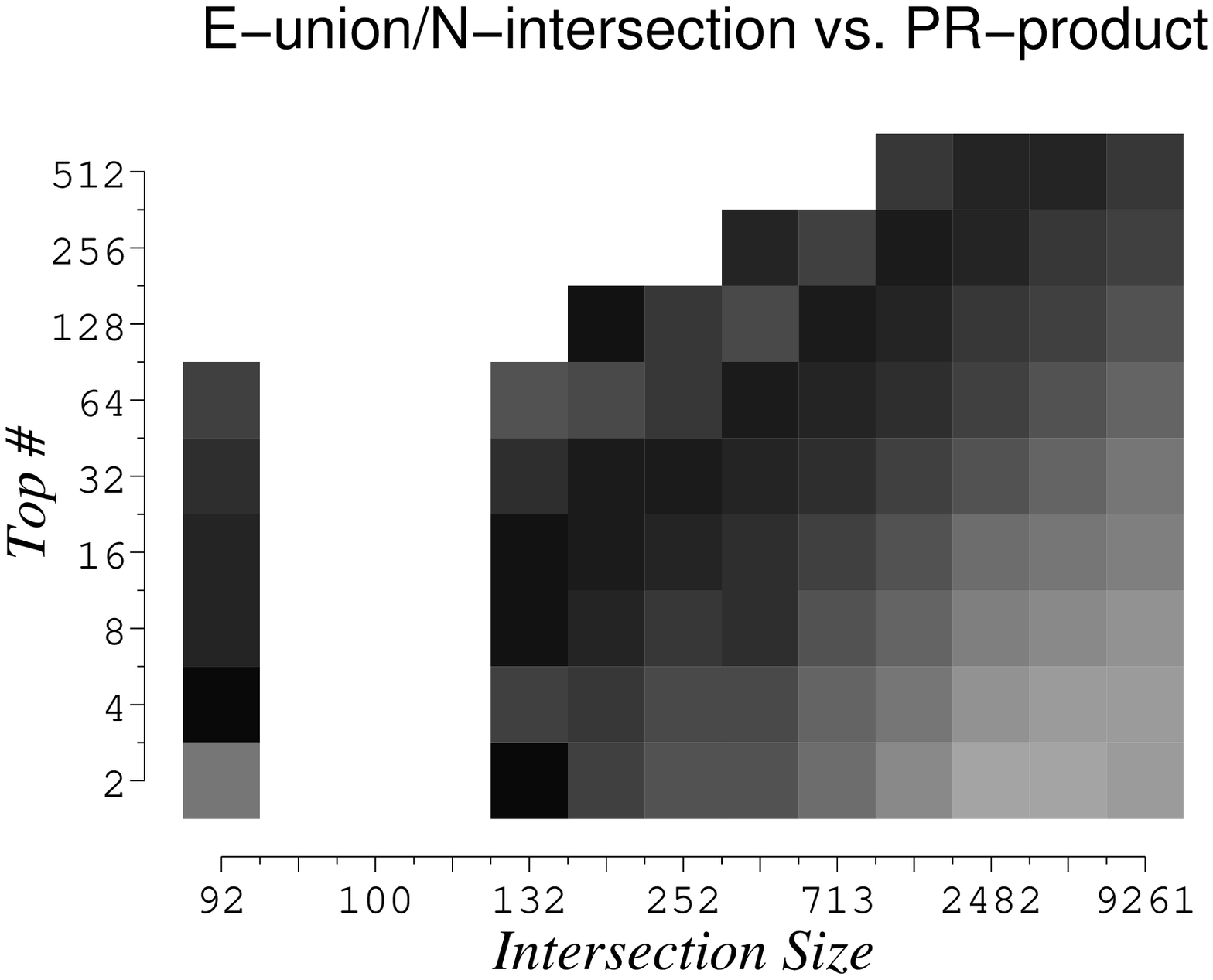}

\includegraphics[height=40mm,angle=0]{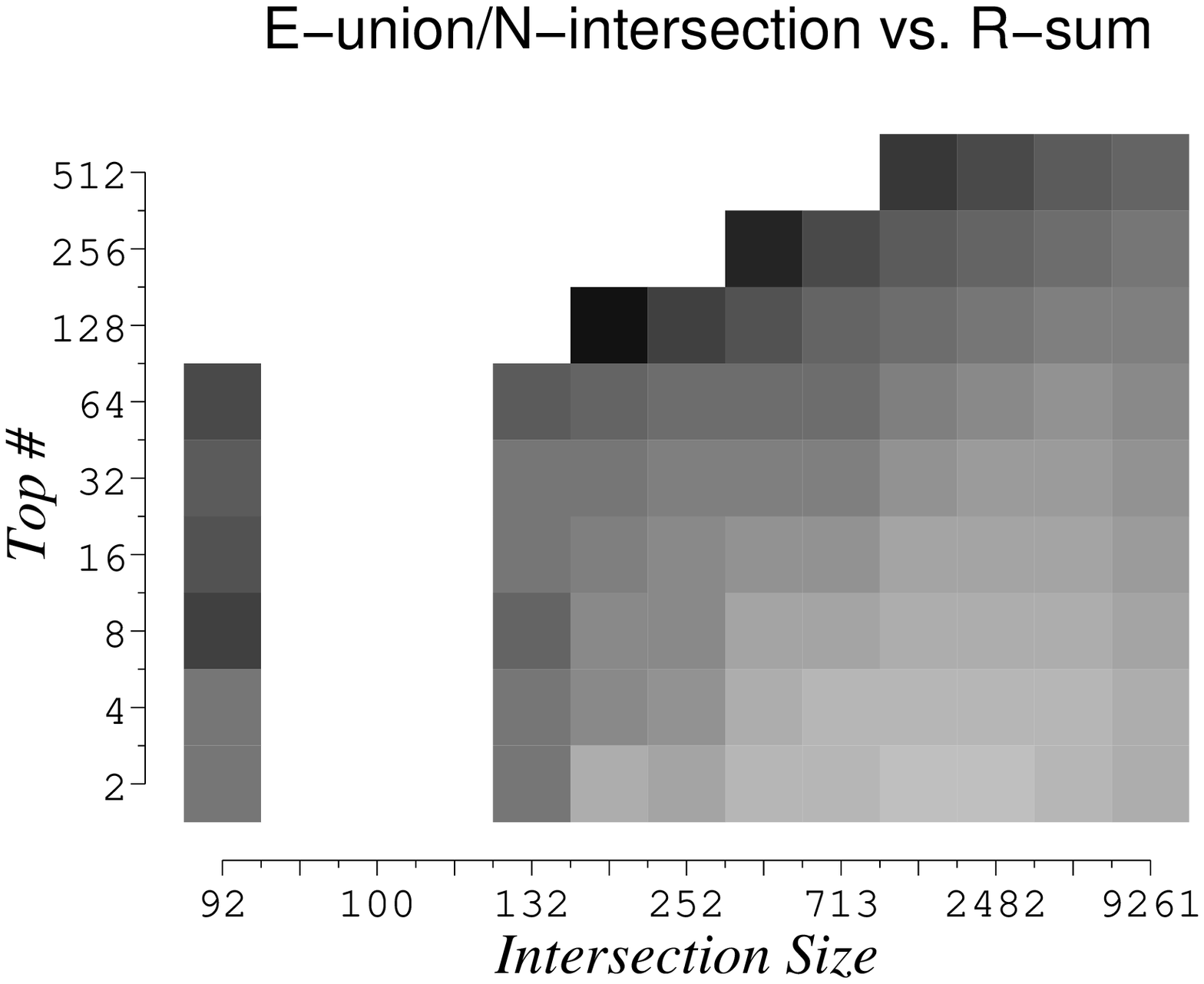}
\includegraphics[height=40mm,angle=0]{bar}
\includegraphics[height=40mm,angle=0]{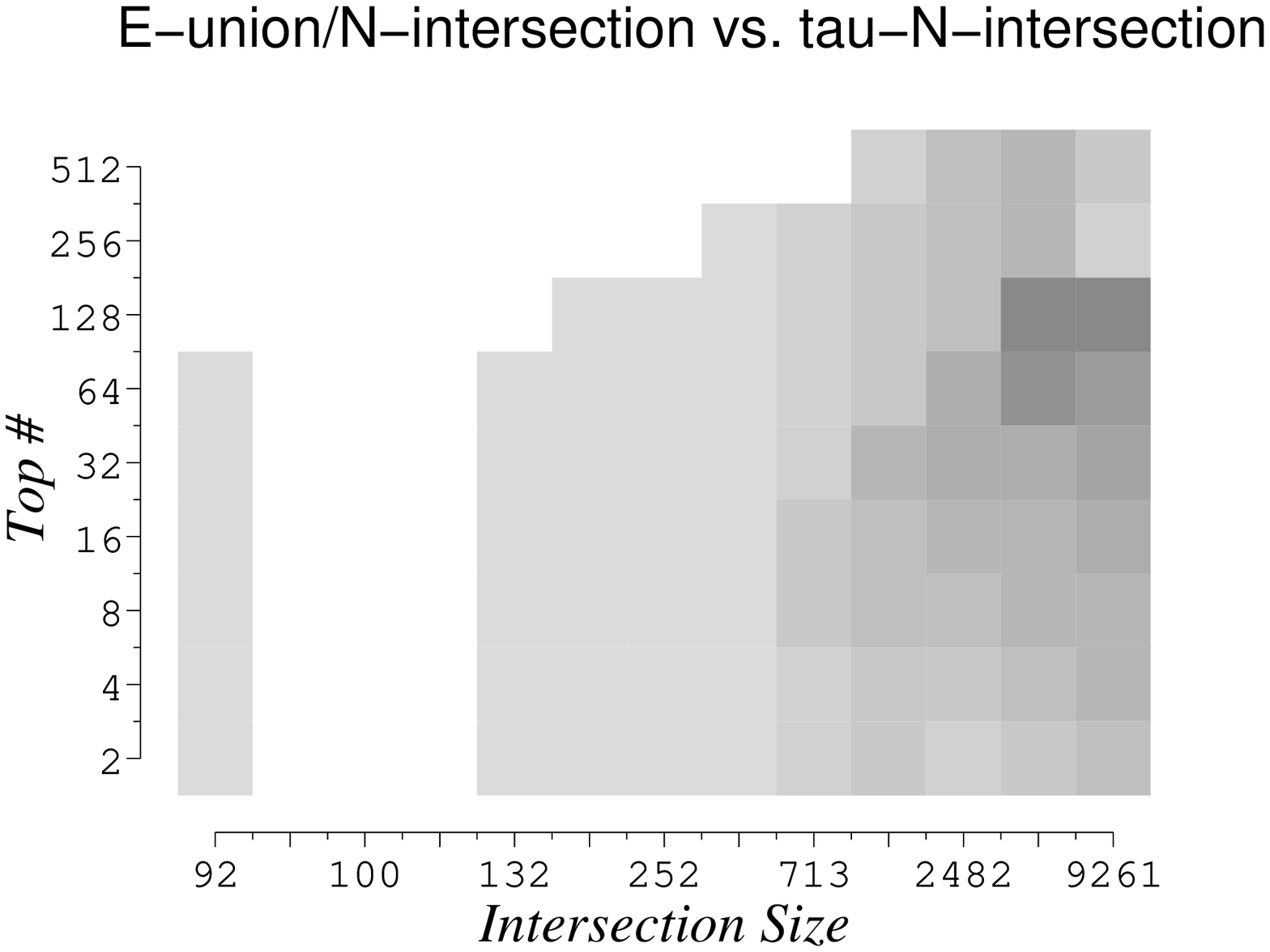}
\end{center}
\caption{Photos network: Average similarity to $E$-union/$N$-intersection}
\label{fig:photos_gray_e-union}
\end{figure}

Table~\ref{tab:photos_comparison} as well as
Figures~\ref{fig:photos_gray_e-intersec}
and~\ref{fig:photos_gray_e-union} summarize the results.

\begin{table}[h]{%\small
\begin{center}
\begin{tabular}{l|ccc}
\multicolumn{4}{c}{{\bf Average similarity to $E$-intersection}}\\
\hline
\hline
Algorithm & \multicolumn{3}{|c}{OSim$|$KSim} \\
     & top 8 & top 16 & top 32 \\
\hline
Single & 0.08$|$0.48 & 0.10$|$0.50 & 0.13$|$0.51 \\
$PR$-product & 0.36$|$0.56 & 0.37$|$0.58 & 0.39$|$0.59 \\
$R$-sum & \bf{0.53}$|$\bf{0.63} & \bf{0.53}$|$\bf{0.64} & \bf{0.52}$|$\bf{0.66} \\
$\tau$-$N$-inters & 0.15$|$0.49 & 0.15$|$0.51 & 0.10$|$0.51 \\
\hline
\end{tabular}

\vspace{10pt}

\begin{tabular}{l|ccc}
\multicolumn{4}{c}{{\bf Average similarity to $E$-union/$N$-intersection}}\\
\hline
\hline
Algorithm & \multicolumn{3}{|c}{OSim$|$KSim} \\
     & top 8 & top 16 & top 32 \\
\hline
Single & 0.31$|$0.53 & 0.34$|$0.55 & 0.39$|$0.56 \\
$PR$-product & \bf{0.72}$|$\bf{0.70} & \bf{0.78}$|$\bf{0.74} & \bf{0.83}$|$\bf{0.79} \\
$R$-sum & 0.35$|$0.54 & 0.42$|$0.56 & 0.50$|$0.59 \\
$\tau$-$N$-inters & 0.13$|$0.49 & 0.12$|$0.51 & 0.09$|$0.51 \\
\hline
\end{tabular}

\caption{Videos network: Comparison of ranking algorithms}
\label{tab:videos_comparison}
\end{center} }
\end{table}

\begin{table}[h]{%\small
\begin{center}
\begin{tabular}{l|ccc}
\multicolumn{4}{c}{{\bf Average similarity to $E$-intersection}}\\
\hline
\hline
Algorithm & \multicolumn{3}{|c}{OSim$|$KSim} \\
     & top 8 & top 16 & top 32 \\
\hline
Single & 0.07$|$0.48 & 0.09$|$0.49 & 0.11$|$0.50 \\
$PR$-product & 0.44$|$0.59 & 0.43$|$0.60 & 0.42$|$0.60 \\
$R$-sum & \bf{0.52}$|$\bf{0.62} & \bf{0.52}$|$\bf{0.63} & \bf{0.52}$|$\bf{0.64} \\
$\tau$-$N$-inters & 0.28$|$0.51 & 0.34$|$0.54 & 0.39$|$0.56 \\
\hline
\end{tabular}

\vspace{10pt}

\begin{tabular}{l|ccc}
\multicolumn{4}{c}{{\bf Average similarity to $E$-union/$N$-intersection}}\\
\hline
\hline
Algorithm & \multicolumn{3}{|c}{OSim$|$KSim} \\
     & top 8 & top 16 & top 32 \\
\hline
Single & 0.17$|$0.50 & 0.21$|$0.51 & 0.27$|$0.53 \\
$PR$-product & \bf{0.50}$|$\bf{0.57} & \bf{0.59}$|$\bf{0.62} & \bf{0.67}$|$\bf{0.66} \\
$R$-sum & 0.28$|$0.52 & 0.32$|$0.54 & 0.38$|$0.56 \\
$\tau$-$N$-inters & 0.19$|$0.50 & 0.22$|$0.52 & 0.26$|$0.53 \\
\hline
\end{tabular}
\caption{Photos network: Comparison of ranking algorithms}
\label{tab:photos_comparison}
\end{center} }
\end{table}

\subsection{Discussion}

\begin{figure}[H!]
\begin{center}
\includegraphics[height=40mm,angle=0]{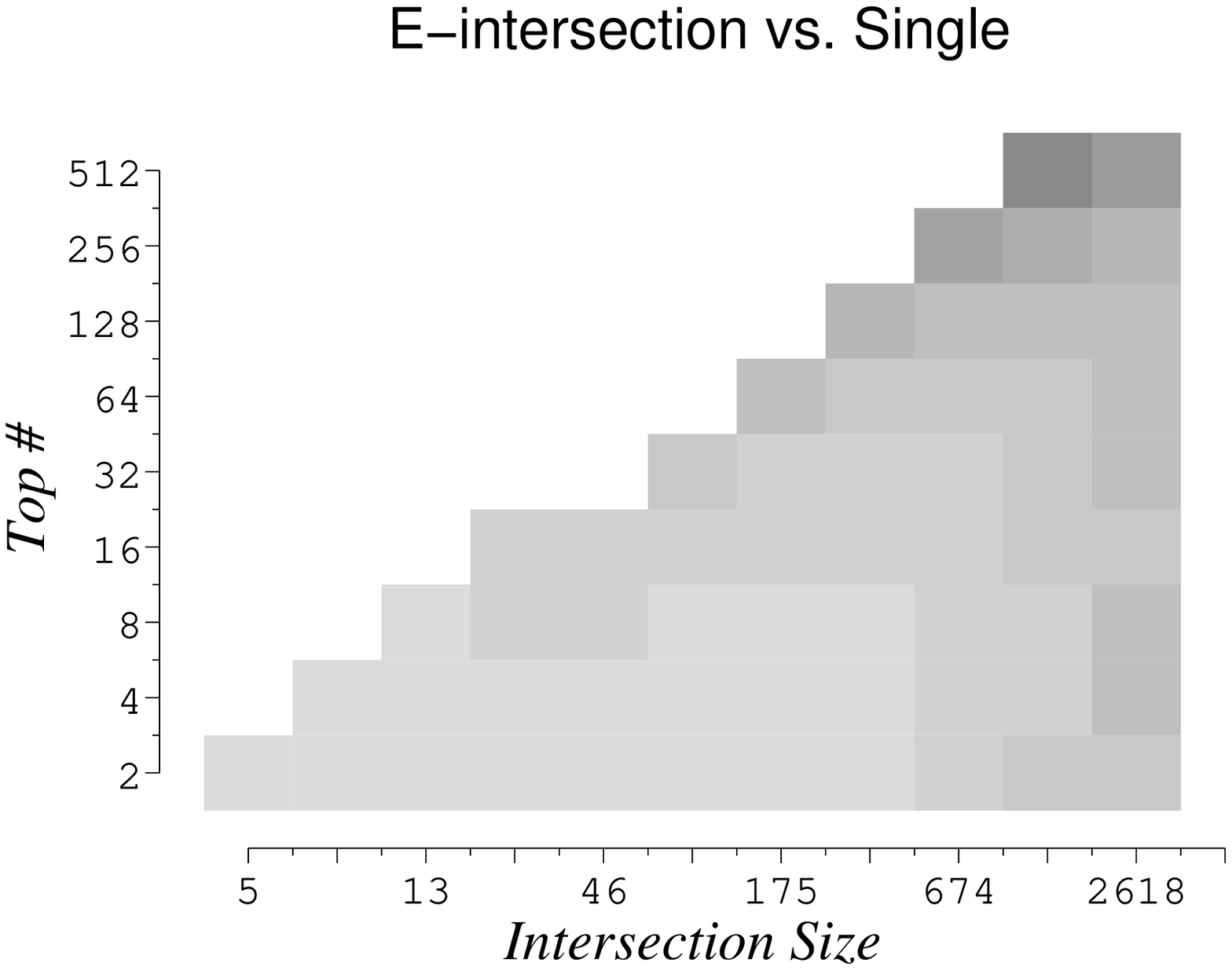}
\includegraphics[height=40mm,angle=0]{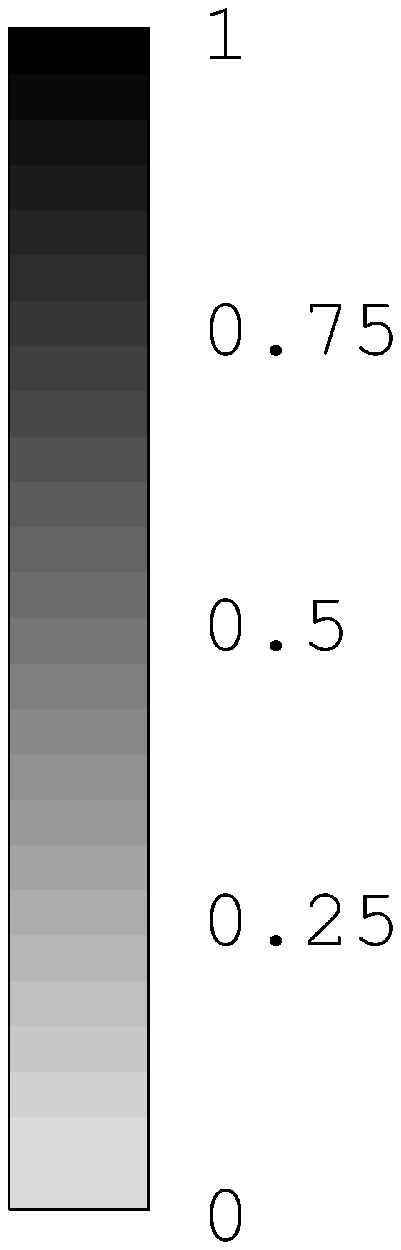}
\includegraphics[height=40mm,angle=0]{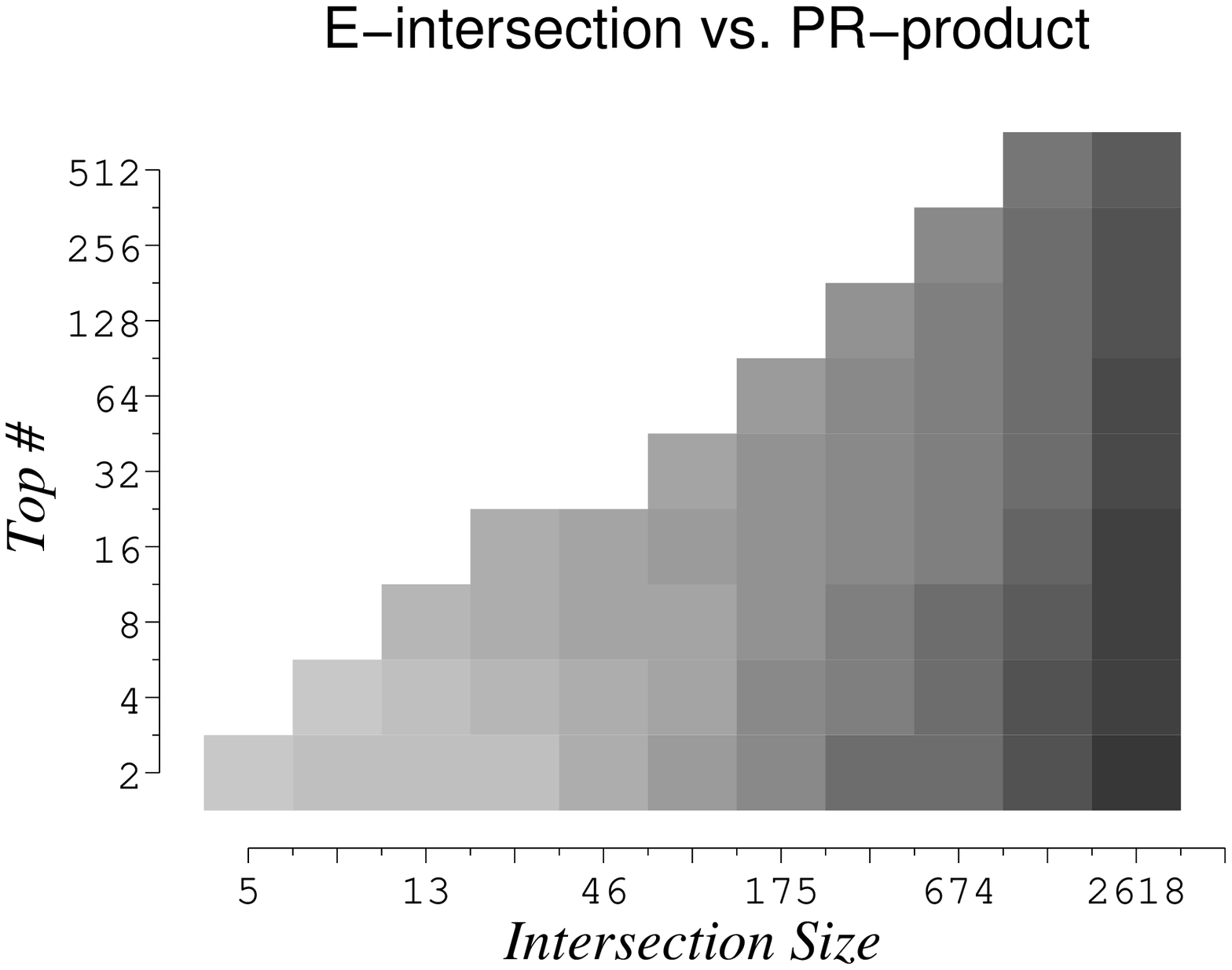}
\includegraphics[height=40mm,angle=0]{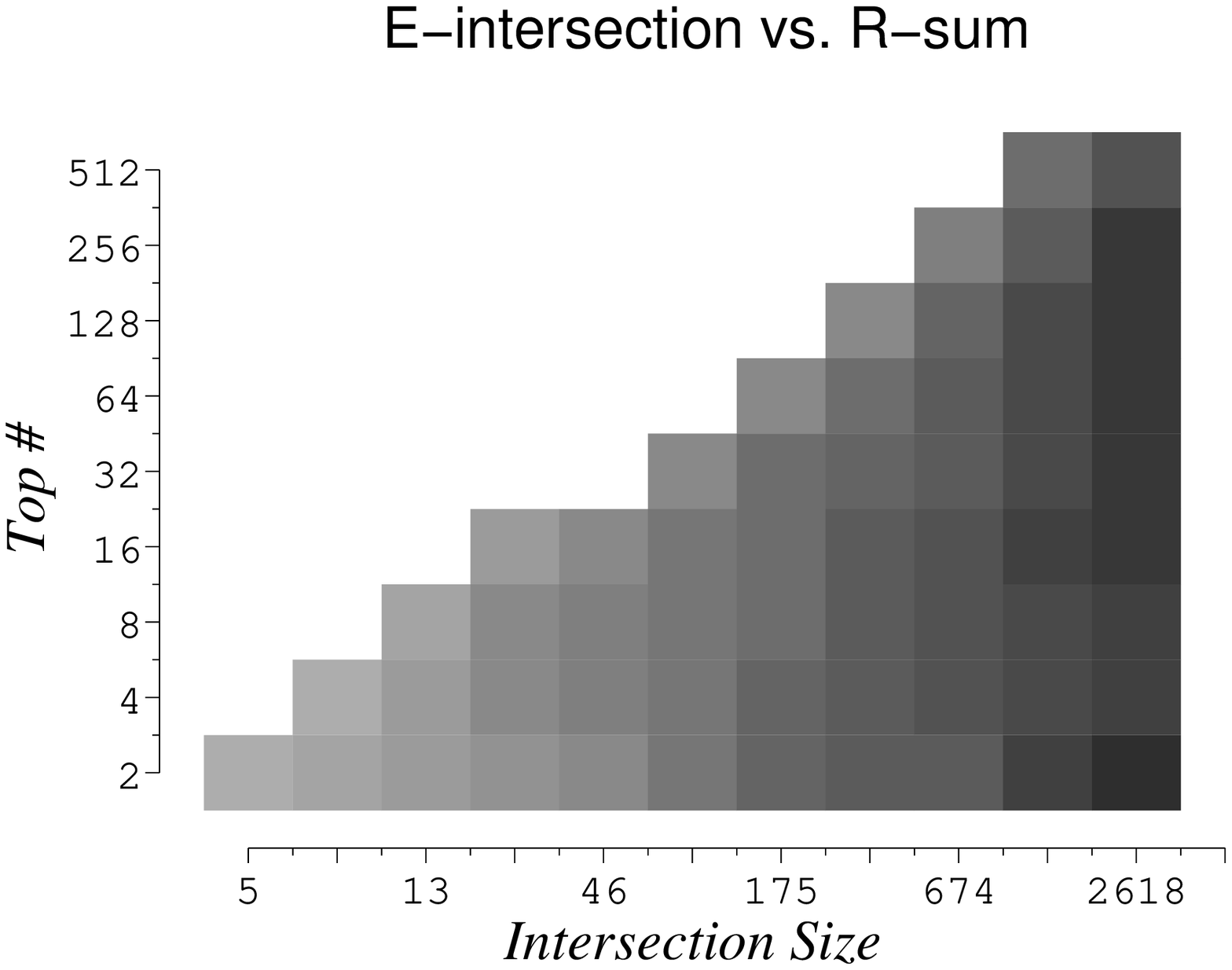}
\includegraphics[height=40mm,angle=0]{bar}
\includegraphics[height=40mm,angle=0]{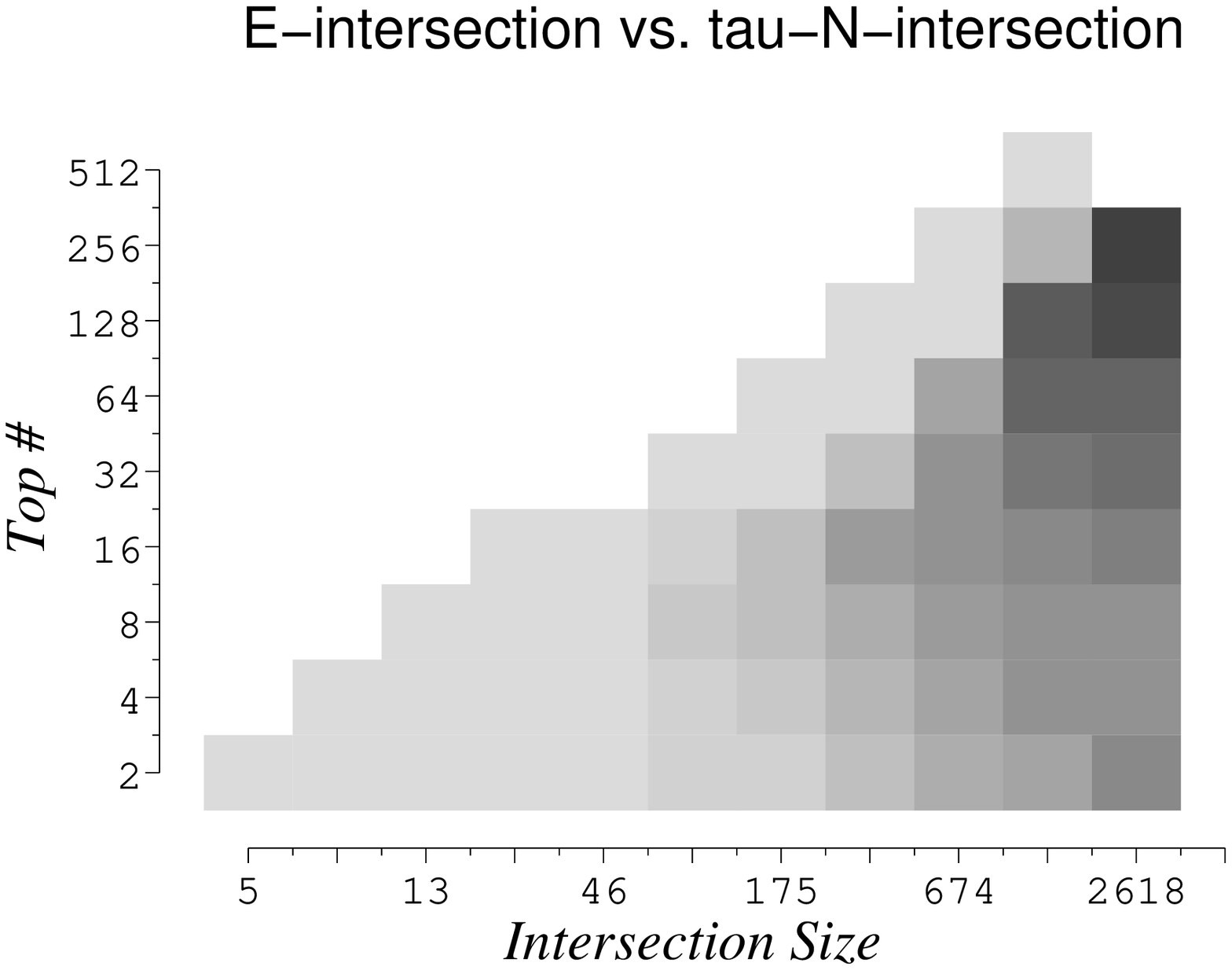}
\end{center}
\caption{Videos network: Average similarity ($OSim$) to $E$-intersection}
\label{fig:videos_gray_e-intersec}
\end{figure}

\begin{figure}[H!]
\begin{center}
\includegraphics[height=40mm,angle=0]{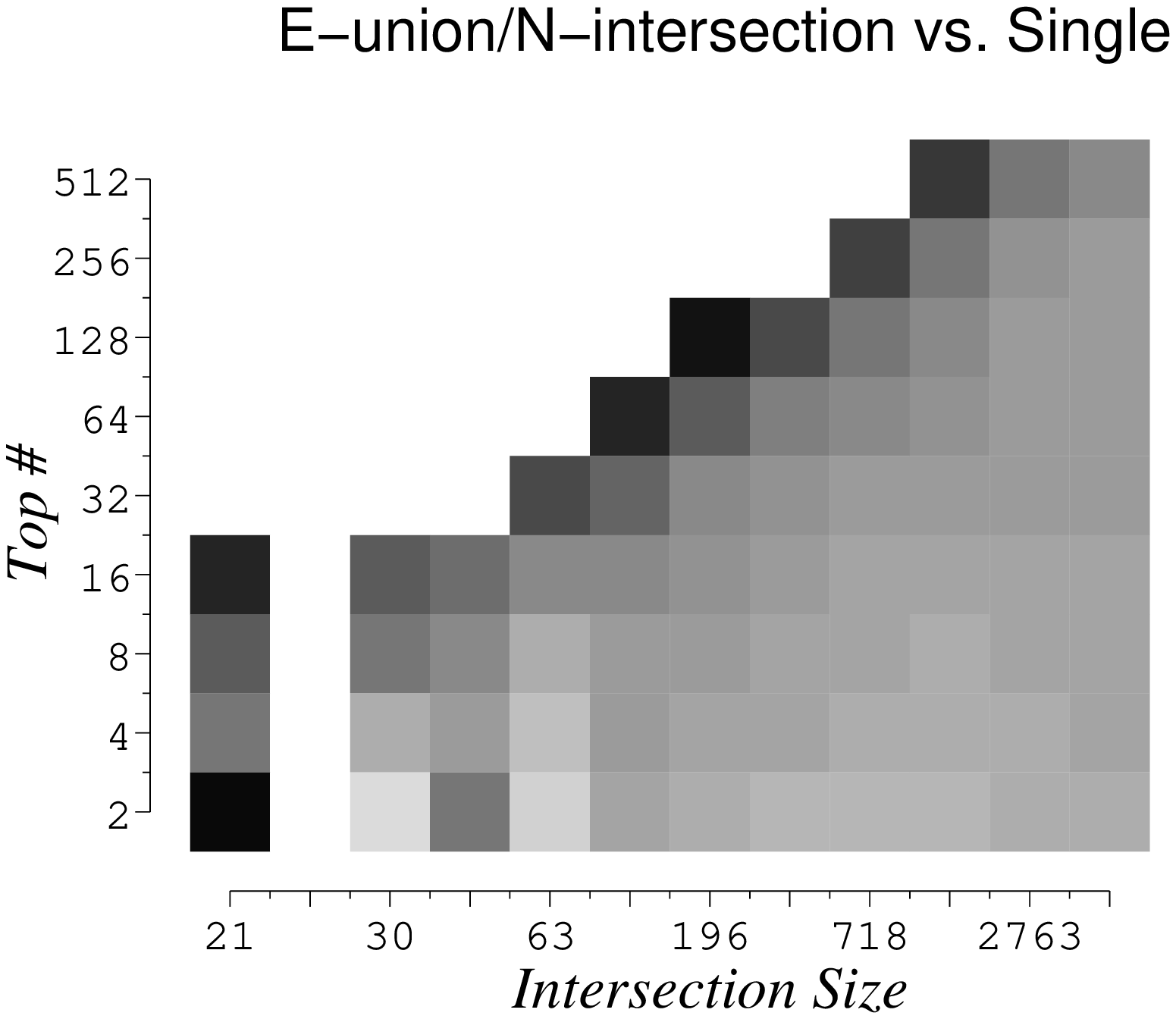}
\includegraphics[height=40mm,angle=0]{bar}
\includegraphics[height=40mm,angle=0]{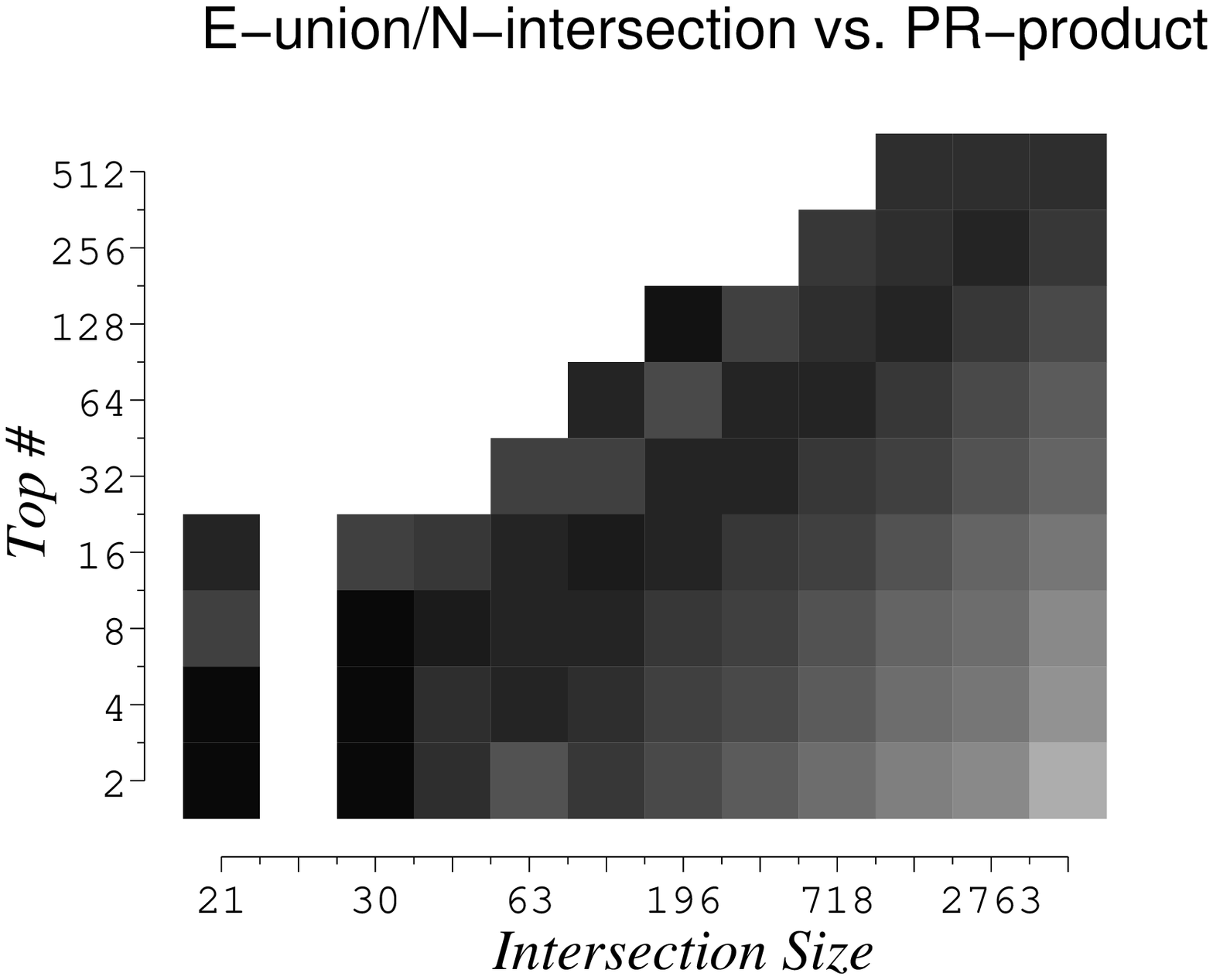}
\includegraphics[height=40mm,angle=0]{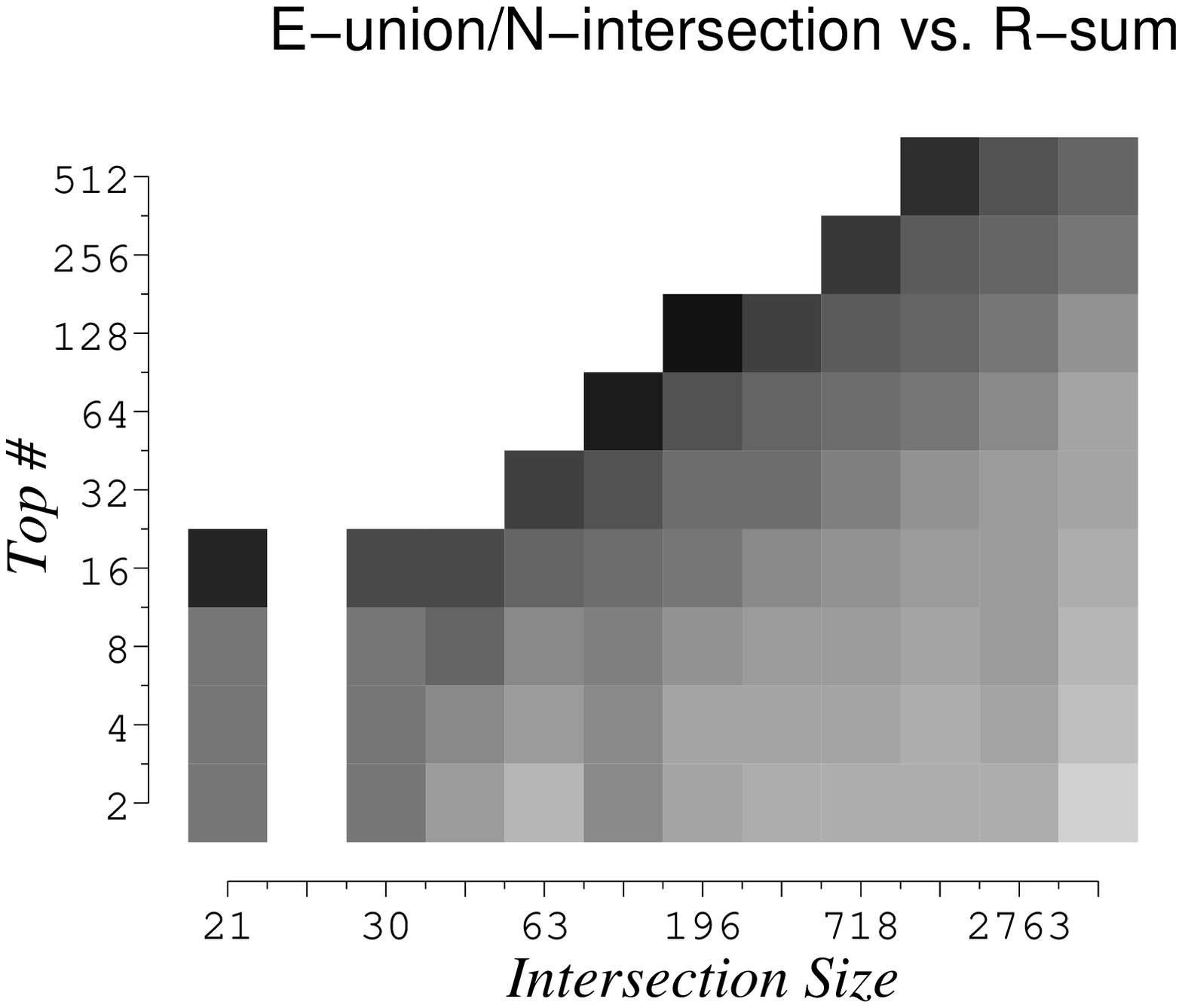}
\includegraphics[height=40mm,angle=0]{bar}
\includegraphics[height=40mm,angle=0]{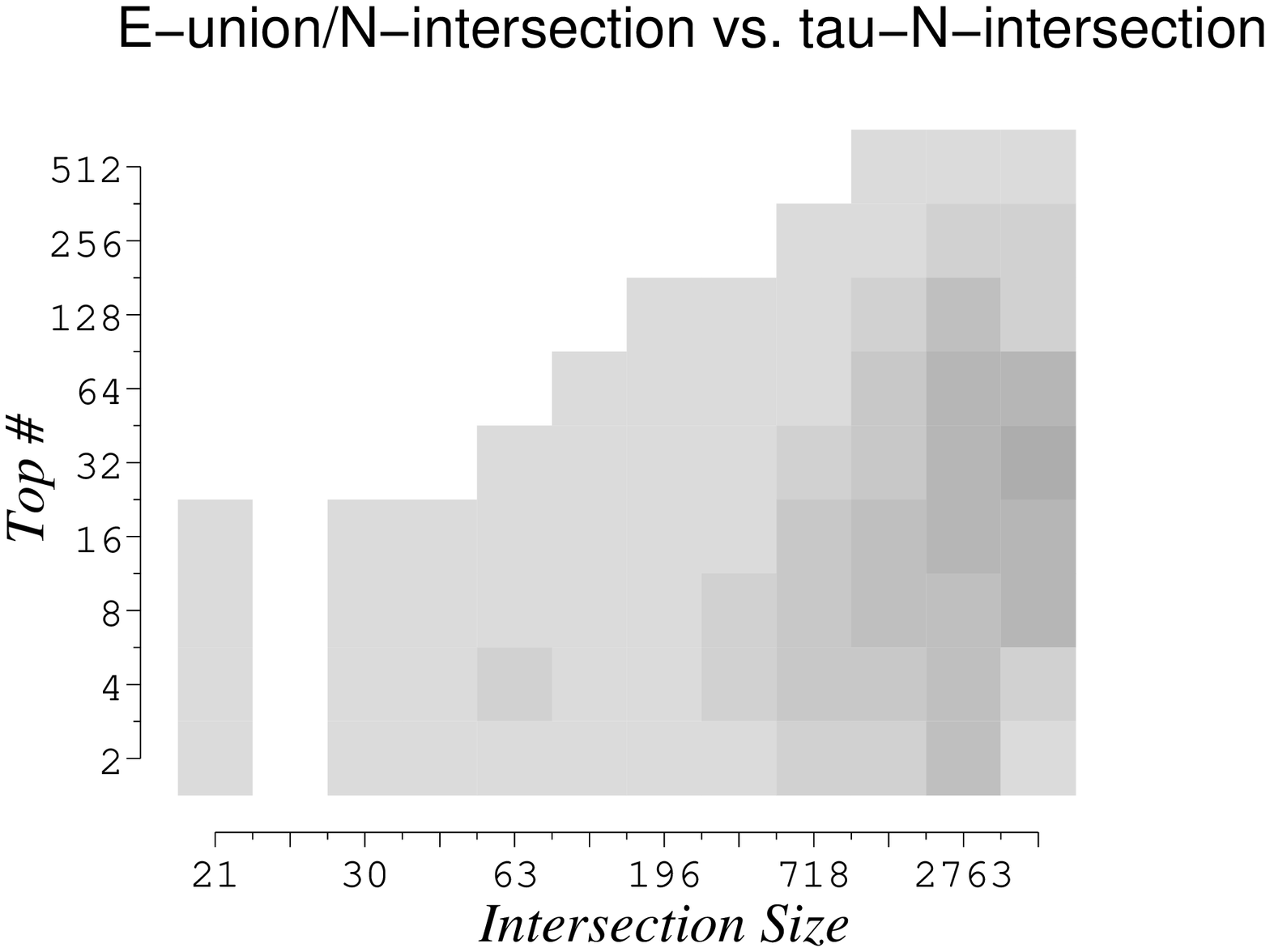}
\end{center}
\caption{Videos network: Average similarity ($OSim$) to $E$-union/$N$-intersection}
\label{fig:videos_gray_e-union}
\end{figure}

As can be appreciated from Tables~\ref{tab:videos_comparison}-\ref{tab:photos_comparison} and Figures \ref{fig:videos_gray_e-intersec}-\ref{fig:videos_gray_e-union}, the Single Ranking algorithm gave the worst results in most cases.

Since the $\tau$-$N$-intersection algorithm is based on retaining only the 500 top-ranked users for each tag, it is natural to observe a worse $OSim$ measure than the other algorithms especially for larger than 500-node intersections. However, this algorithm gives worse results even for smaller intersections. This fact is explained by the relevance of a large number of recommendations of low-ranked users when computing the PageRank in both the $E$-intersection and the $E$-union/$N$-intersection cases. Also note that the $\tau$-$N$-intersection approach gave better results on the photo network than in the video network. A possible cause is the assortativeness of photo network (see Figure~\ref{fig:netinfo_inoutcorrel} and Section~\ref{subsec:network-analysis}). Indeed, since assortativeness implies that users with many recommendations are preferentially recommended by users with also many recommendations, the relevance of low-ranked users in the computation of the centrality measure is lower.

There is a remarkable improvement using algorithm $R$-sum compared to the other merging algorithms when considering the similarity to the $E$-intersection standard on both networks. Also, the best merging algorithm for the similarity with the second standard $E$-union/$N$-intersection is $PR$-product merging algorithm.

\section{Summary}
\label{sec:conclusions}

We have proposed different algorithms for merging faceted-rankings of users in collaborative tagging systems which gave results comparable to those of two reasonable standards. We have also analyzed the scalability of this approach.

A prototypic application called \emph{Egg-O-Matic} is available online~\cite{eggomatic} including ranking merging $R$-sum to approximate the $E$-intersection ranking, in a mode called \emph{``all tags, same content''}, and including the ranking merging we called $PR$-product to approximate the $E$-union/$N$-intersection ranking, in a mode called \emph{``all tags, any content''}.

Another step that can be taken to reduce tag-dimensionality is clustering to agglomerate them. This work also opens the path for a more complex comparison of reputations, for example by integrating the best positions of a user even if the tags involved are not related (\emph{disjunctive} queries) in order to summarize the relevance of a user generating content on the web.
It is also possible to extend the algorithms in Section~\ref{sec:definitions_and_algorithms} to merge of rankings generated from different systems (\emph{cross-system ranking}) looking to obtain a ranking of users using multiple collaborative tagging systems.

\bibliographystyle{alpha}%\bibliographystyle{latex8}
\bibliography{articulo}

\newpage

\onecolumn

\end{document}